\documentclass[proceedings]{JHEP3}
\usepackage{amsfonts}
\usepackage{amsmath}
\usepackage{epsfig}

\newbox\mybox

\newcommand\fverb{\setbox\mybox=\hbox\bgroup\verb}
\newcommand\fverbdo{\egroup\medskip\noindent\fbox{\unhbox\mybox}\ }
\newcommand\fverbit{\egroup\item[\fbox{\unhbox\mybox}]}
\conference{Bohmian quantum trajectories from coherent states}
\abstract{We find that real and complex Bohmian quantum trajectories resulting 
from well-localized Klauder coherent states 
in the quasi-Poissonian regime possess qualitatively the same type of trajectories as those obtained 
from a purely classical analysis of the corresponding Hamilton-Jacobi equation. In the complex
cases treated the quantum potential results to a constant, such that the agreement is exact.
For the real cases we provide 
conjectures for analytical solutions for the trajectories as well as the corresponding quantum potentials. 
The overall qualitative behaviour is governed by the Mandel parameter determining the regime in which 
the wavefuntions evolve as soliton like structures. We demonstrate these features explicitly for 
the harmonic oscillator and the P\"{o}schl-Teller potential.}

\title{Bohmian quantum trajectories from coherent states}
\author{Sanjib Dey and Andreas Fring \\
Department of Mathematical Science, City University London,\\
Northampton Square, London EC1V 0HB, UK\\
E-mail: sanjib.dey.1@city.ac.uk, a.fring@city.ac.uk}

\input{tcilatex}
\begin{document}

\section{Introduction}

Bohmian mechanics was originally proposed sixty years ago \cite{Bohm:1951xw}
to address some of the difficulties present in the standard formulation of
quantum mechanics based on the Copenhagen interpretation and its aim was to
provide an alternative ontological view. Its central purpose is to avoid the
need for the collapse of the wavefunction and instead provide a trajectory
based scheme allowing for a causal interpretation. While this metaphysical
discussion is still ongoing and is in parts very controversial \cite%
{BohmHiley,Holland,HileyB}, it needs to be stressed that Bohmian mechanics
leads to the same predictions of measurable quantities as the orthodox
framework. Here we will leave the interpretational issues aside and build on
the fact that the Bohmian formulation of quantum mechanics has undoubtedly
proven to be a successful technical tool for the study of some concrete
physical scenarios. For instance, it has been applied successfully to study
of photodissociation problems \cite{SAR}, tunneling processes \cite%
{PhysRevLett.82.5190}, atom diffraction by surfaces \cite%
{PhysRevB.61.7743,WDH,GSMM} and high harmonic generation \cite{CarlaBohm}.
Whereas these applications are mainly based on an analysis of real valued
quantum trajectories, more recently there has also been the suggestion \cite%
{PhysRevLett.50.3,PhysRevD.28.2491} for a formulation of Bohmian mechanics
based on complex trajectories. We will discuss here both versions, but it is
this latter formulation on which we will place our main focus and which will
be the main subject of our investigations in this manuscript.

Independently from the above suggestions, an alternative perspective on
complex classical mechanics has recently emerged out of the study of complex
quantum mechanical Hamiltonians. It is by now well accepted that a large
class of such systems constitute well-defined self-consistent descriptions
of physical systems \cite{Bender:1998ke,Benderrev,Alirev} with real energy
eigenvalue spectra and unitary time-evolution. The dynamics of many
classical models has been investigated, for instance complex extensions of
standard one particle systems \cite{Nana,Bender:2006tz,Bender:2010eg},
non-Hamiltonian dynamical systems \cite{Bender:2007pr}, chaotic systems \cite%
{Bender:2008qe} or deformations of many-particle systems such as
Calogero-Moser-Sutherland models \cite{AF,AFZ,Assis:2009gt,FringSmith}. From
those studies conclusions were drawn for example with regard to tunneling
behaviour \cite{Bender:2010eg} or the existence of band structures \cite%
{Benderbands}. It was also shown \cite{CFB} that complex solutions to a
large class of complex quantum mechanical systems arise as special cases
from the study of Korteweg-deVries type of field equations, see \cite{AFrev}
for a review on new models obtained from deformations of integrable systems.

It is therefore natural to compare these two formulations and address the
question of whether they are equivalent in some regime. We will demonstrate
here that this is indeed the case, since the complex Bohmian quantum
mechanics based on so-called Klauder coherent states \cite%
{Klauder0,Klauder01,Klauder1,Klauder2} in the quasi-Poissonian regime is
identical to a purely classical study of the Hamilton-Jacobi equations.

Our manuscript is organized as follows: In section two we recall the basic
equations of Bohmian mechanics in the real as well as the complex case
together with the main features of Klauder coherent states. In section 3 we
compare in both cases the trajectories resulting from standard Gaussian
wavepackets and those resulting from Klauder coherent states with a purely
classical treatment. For the latter case we verify the validity of a
conjectured formula for the trajectories. In section 4 we discuss the same
scenario for the P\"{o}schl-Teller potential. Our conclusions are stated in
section 5

\section{Real and complex Bohmian mechanics, Klauder coherent states}

Let us briefly recall the key equations of Bohmian mechanics for reference
purposes and also to establish our conventions and notations. The starting
point for the construction of the Bohmian quantum trajectories is usually a
solution of the time-dependent Schr\"{o}dinger equation involving a
potential $V(x)$ 
\begin{equation}
i\hbar \frac{\partial \psi (x,t)}{\partial t}=-\frac{\hbar ^{2}}{2m}\frac{%
\partial ^{2}\psi (x,t)}{\partial x^{2}}+V(x)\psi (x,t)\text{.}  \label{SE}
\end{equation}%
The two variants leading either to real or complex trajectories are
distinguished by different parameterizations of the wavefunctions.

\subsection{The real variant}

The version based on real valued trajectories results from the WKB-polar
decomposition%
\begin{equation}
\psi (x,t)=R(x,t)e^{i/\hbar S(x,t)},\text{\qquad with \ \ }R(x,t),S(x,t)\in 
\mathbb{R}\text{.}  \label{phi1}
\end{equation}%
Upon the substitution of (\ref{phi1}) into (\ref{SE}) the real and imaginary
part are identified as%
\begin{equation}
S_{t}+\frac{(S_{x})^{2}}{2m}+V(x)-\frac{\hbar ^{2}}{2m}\frac{R_{xx}}{R}%
=0,\qquad \text{and\qquad }mR_{t}+R_{x}S_{x}+\frac{1}{2}RS_{xx}=0,
\label{r1}
\end{equation}%
usually referred to as the quantum Hamilton-Jacobi equation and the
continuity equation, respectively. When considering these equations from a
classical point of view, the second term in the first equation of (\ref{r1})
is interpreted as the kinetic energy such that the real velocity $v(t)$ and
the last term, the so-called quantum potential $Q(x,t)$, result to%
\begin{equation}
mv(x,t)=S_{x}=\frac{\hbar }{2i}\left[ \frac{\psi ^{\ast }\psi _{x}-\psi \psi
_{x}^{\ast }}{\psi ^{\ast }\psi }\right] ,~~Q(x,t)=-\frac{\hbar ^{2}}{2m}%
\frac{R_{xx}}{R}=\frac{\hbar ^{2}}{4m}\left[ \frac{\left( \psi ^{\ast }\psi
\right) _{x}^{2}}{2\left( \psi ^{\ast }\psi \right) ^{2}}-\frac{\left( \psi
^{\ast }\psi \right) _{xx}}{\psi ^{\ast }\psi }\right] ,  \label{real}
\end{equation}%
respectively. The corresponding time-dependent effective potential is
therefore $V_{\text{eff}}(x,t)=V(x)+Q(x,t)$. Then one has two options to
compute quantum trajectories. One can either solve directly the first
equation in (\ref{real}) for $x(t)$ or employ the effective potential $V_{%
\text{eff}}$ solving $m\ddot{x}=-\partial V_{\text{eff}}/\partial x$
instead. Due to the different order of the differential equations to be
solved, we have then either one or two free parameter available. Thus for
the two possibilities to coincide the initial momentum is usually not free
of choice, but the initial position $x(t=0)=x_{0}$ is the only further
input. The connection to the standard quantum mechanical description is then
achieved by computing expectation values from an ensemble of $n$
trajectories, e.g. $\left\langle x(t)\right\rangle
_{n}=1/n\dsum\nolimits_{i=1}^{n}x_{i}(t)$.

\subsection{The complex variant}

In contrast, the version based on complex trajectories is computed from a
parameterization of the form 
\begin{equation}
\psi (x,t)=e^{i/\hbar \tilde{S}(x,t)},\text{\qquad with }\tilde{S}(x,t)\in 
\mathbb{C}.  \label{phi2}
\end{equation}%
The substitution of (\ref{phi2}) into (\ref{SE}) yields the single equation%
\begin{equation}
\tilde{S}_{t}+\frac{(\tilde{S}_{x})^{2}}{2m}+V(x)-\frac{i\hbar }{2m}\tilde{S}%
_{xx}=0.  \label{c1}
\end{equation}%
Interpreting this equation in a similar way as in the previous subsection,
but now as a complex quantum Hamilton-Jacobi equation, the second term in (%
\ref{c1}) yields a complex velocity and the last term becomes a complex
quantum potential 
\begin{equation}
m\tilde{v}(x,t)=\hat{S}_{x}=\frac{\hbar }{i}\frac{\psi _{x}}{\psi },~\ ~%
\tilde{Q}(x,t)=-\frac{i\hbar }{2m}\tilde{S}_{xx}=-\frac{\hbar ^{2}}{2m}\left[
\frac{\psi _{xx}}{\psi }-\frac{\psi _{x}^{2}}{\psi ^{2}}\right] .
\label{complex}
\end{equation}%
The corresponding time-dependent effective potential is now $\tilde{V}_{%
\text{eff}}(x,t)=V(x)+\tilde{Q}(x,t)$. Once again one has two options to
compute quantum trajectories, either solving the first equation in (\ref%
{complex}) for $x(t)$, which is, however, now a complex variable.
Alternatively, we may also view the effective Hamiltonian $H_{\text{eff}%
}=p^{2}/2m+\tilde{V}_{\text{eff}}(x,t)=H_{r}+iH_{i}$ in its own right and
simply compute the equations of motion directly from 
\begin{eqnarray}
\dot{x}_{r} &=&\frac{1}{2}\left( \frac{\partial H_{r}}{\partial p_{r}}+\frac{%
\partial H_{i}}{\partial p_{i}}\right) ,\qquad \dot{x}_{i}=\frac{1}{2}\left( 
\frac{\partial H_{i}}{\partial p_{r}}-\frac{\partial H_{r}}{\partial p_{i}}%
\right) ,  \label{H1} \\
\dot{p}_{r} &=&-\frac{1}{2}\left( \frac{\partial H_{r}}{\partial x_{r}}+%
\frac{\partial H_{i}}{\partial x_{i}}\right) ,~~~~~~\dot{p}_{i}=\frac{1}{2}%
\left( \frac{\partial H_{r}}{\partial x_{i}}-\frac{\partial H_{i}}{\partial
x_{r}}\right) ,  \label{H2}
\end{eqnarray}%
where we use the notations $x=x_{r}+ix_{i}$ and $p=p_{r}+ip_{i}$ with $x_{r}$%
, $x_{i}$, $p_{r}$, $p_{i}\in \mathbb{R}$. For the complex case the relation
to the conventional quantum mechanical picture is less well established
although some versions have been suggested to extract real expectation
values, e.g. based on taking time-averaged mean values \cite{ComplYang},
seeking for isochrones \cite{Goldfarb,ChouW} or using imaginary part of the
velocity field of particles on the real axis \cite{MVJohn2}.

We will here evaluate the expressions for the velocity, the quantum
potential and the resulting trajectories for two solvable potentials
commencing with different choices of solutions $\psi $. Our particular focus
is here on commencing from coherent states and for that reason we state
their main properties.

\subsection{Klauder coherent states}

For Hermitian Hamiltonians $\mathcal{H}$, with discrete bounded below and
nondegenerate eigenspectrum $E_{n}=\omega e_{n}$ and orthonormal eigenstates 
$\left\vert \phi _{n}\right\rangle $ the Klauder coherent states \cite%
{Klauder0,Klauder01,Klauder1,Klauder2} are defined in general as 
\begin{equation}
\psi _{J}(x,t):=\frac{1}{\mathcal{N}(J)}\sum\limits_{n=0}^{\infty }\frac{%
J^{n/2}\exp (-i\omega te_{n})}{\sqrt{\rho _{n}}}\phi _{n}(x),\text{\qquad }%
J\in \mathbb{R}_{0}^{+}.  \label{GK}
\end{equation}%
The probability distribution and the normalization constant are given by $%
\rho _{n}:=\dprod\nolimits_{k=1}^{n}e_{k}$ and $\mathcal{N}%
^{2}(J):=\dsum\nolimits_{k=0}^{\infty }J^{k}/\rho _{k}$, respectively. To
allow for a more compact notation we adopt the usual convention $\rho _{0}=1$
throughout the manuscript. The key properties of these states are their
continuity in time and the variable $J$, the fact that they provide a
resolution of the identity and that they are temporarily stable satisfying
the action angle identity $\left\langle J,\omega t\right\vert \mathcal{H}%
\left\vert J,\omega t\right\rangle =\hbar \omega J$.

Evidently the expression for $\psi _{J}(x,t)$ is only meaningful when the
wave packet is properly localized, i.e. the absolute value squared of the
weighting function $c_{n}(J)=J^{n/2}/\mathcal{N}(J)\sqrt{\rho _{n}}$ needs
to be peaked about some mean value $\left\langle n\right\rangle =2Jd\ln 
\mathcal{N}(J)/dJ$. The deviation from a Poissonian distribution is captured
in the so-called Mandel parameter \cite{Mandel} defined as%
\begin{equation}
Q:=\frac{\Delta n^{2}}{\left\langle n\right\rangle }-1=J\frac{d}{dJ}\ln 
\frac{d}{dJ}\ln \mathcal{N}^{2}(J),  \label{Mandel}
\end{equation}%
with dispersion $\Delta n^{2}=\left\langle n^{2}\right\rangle -\left\langle
n\right\rangle ^{2}$. Here the case $Q=0$ is a pure Poisson distribution
with $Q<0$ and $Q>0$ corresponding to sub-Poisson and super-Poisson
distributions, respectively. We refer to distributions with $\left\vert
Q\right\vert \ll 1$ as quasi-Poissonian.

\section{The harmonic oscillator}

The harmonic oscillator 
\begin{equation}
\mathcal{H}_{\text{ho}}=\frac{p^{2}}{2m}+\frac{1}{2}m\omega ^{2}x^{2},
\label{VHO}
\end{equation}%
constitutes a very instructive example on which many of the basic features
can be understood. We will therefore take it as a starting point. Many
results may already be found in the literature, but for completeness we also
report them here together with some new findings.

\subsection{Real case}

As reported for instance by Holland \cite{Holland}, using the two formulae
in (\ref{real}) it is easy to see that for any stationary state $\psi
_{n}(x,t)=\phi _{n}(x)e^{-iE_{n}t/\hbar }$, with $\phi _{n}(x)$ being a
solution of the stationary Schr\"{o}dinger, the velocity in (\ref{real})
results to $v(t)=0$. This is compatible with the values obtained from the
use of the quantum potential $Q(x)=E_{n}-V(x)$, because this corresponds to
a classical motion in a constant effective potential $V_{\text{eff}%
}(x,t)=E_{n}$. This is of course qualitatively very far removed from our
original potential (\ref{VHO}), such that classical trajectories obtained
from $\mathcal{H}_{\text{ho}}$ and its effective version are fundamentally
of qualitatively different nature.

Instead we would expect, that when starting from coherent states we end up
with a behaviour much closer to the classical behaviour resulting from the
original Hamiltonian. By direct computation shown in \cite{Holland}, using
the standard Gaussian wavepackets of the form%
\begin{equation}
\psi _{c}(x,t)=\left( \frac{m\omega }{\hbar \pi }\right) ^{1/4}e^{-\frac{%
m\omega }{2\hbar }(x-a\cos \omega t)^{2}-\frac{i}{2}\left[ \omega t+\frac{%
m\omega }{\hbar }\left( 2xa\sin \omega t-\frac{1}{2}a^{2}\sin 2\omega
t\right) \right] },  \label{fc}
\end{equation}%
to compute the above quantities with $a$ being the centre of the wavepacket
at $t=0$, one obtains from (\ref{real}) equations for the velocity and the
quantum potential as%
\begin{equation}
v(x,t)=-a\omega \sin \omega t,\qquad \text{and\qquad }~Q(x,t)=\frac{\hbar
\omega }{2}-\frac{1}{2}m\omega ^{2}(x-a\cos \omega t)^{2}.  \label{vQ}
\end{equation}%
Solving now the first equation in (\ref{vQ}) with $dx/dt=v(x,t)$ for $%
x(t)=a(\cos \omega t-1)+x_{0}$ with initial condition $x(0)=x_{0}$, we may
construct the corresponding potential from $m\ddot{x}=-\partial V/\partial x$%
. The result is compatible with the effective potential obtained from $%
Q(x,t)+V(x)$ when replacing the explicit time dependent terms with
expressions in $x(t)$. Alternatively, from the effective potential 
\begin{equation}
V_{\text{eff}}[x(t)]=\frac{1}{2}m\omega ^{2}(x(t)-x_{0}-a)^{2}+\frac{\hbar
\omega }{2},
\end{equation}%
Newton's equation will give the above solution for $x(t)$. Thus for the
states $\psi _{c}(x,t)$ the Bohmian trajectories for the harmonic oscillator
potential $\mathcal{H}_{\text{ho}}$ are indeed the same as those resulting
from the motion in a classical harmonic oscillator potential.

What has not been analyzed this far is the use the more general Klauder
coherent states (\ref{GK}) as input into the evaluation of the Bohmian
trajectories and corresponding quantum potential $Q(x,t)$. Since we have $%
e_{n}=n$ for the case at hand, the probability distribution and
normalization constant are computed to $\rho _{n}=n!$ and $\mathcal{N}%
(J)=e^{J/2}$, respectively. The solution for the stationary Schr\"{o}dinger
equation is well known to be the normalized wavefunction $\phi _{n}(x)=(%
\frac{m\omega }{\pi \hbar })^{1/4}\exp \left( -\frac{mx^{2}\omega }{2\hbar }%
\right) H_{n}\left( x/\sqrt{\frac{\hbar }{m\omega }}\right) /\sqrt{2^{n}n!}$
with $H_{n}\left( x\right) $ denoting Hermite polynomials. The Mandel
parameter $Q$ in (\ref{Mandel}) always equals zero independently of $J$,
such that we are always dealing with a Poissonian distribution.

\begin{figure}[h]
\centering   \includegraphics[width=7.5cm,height=6.0cm]{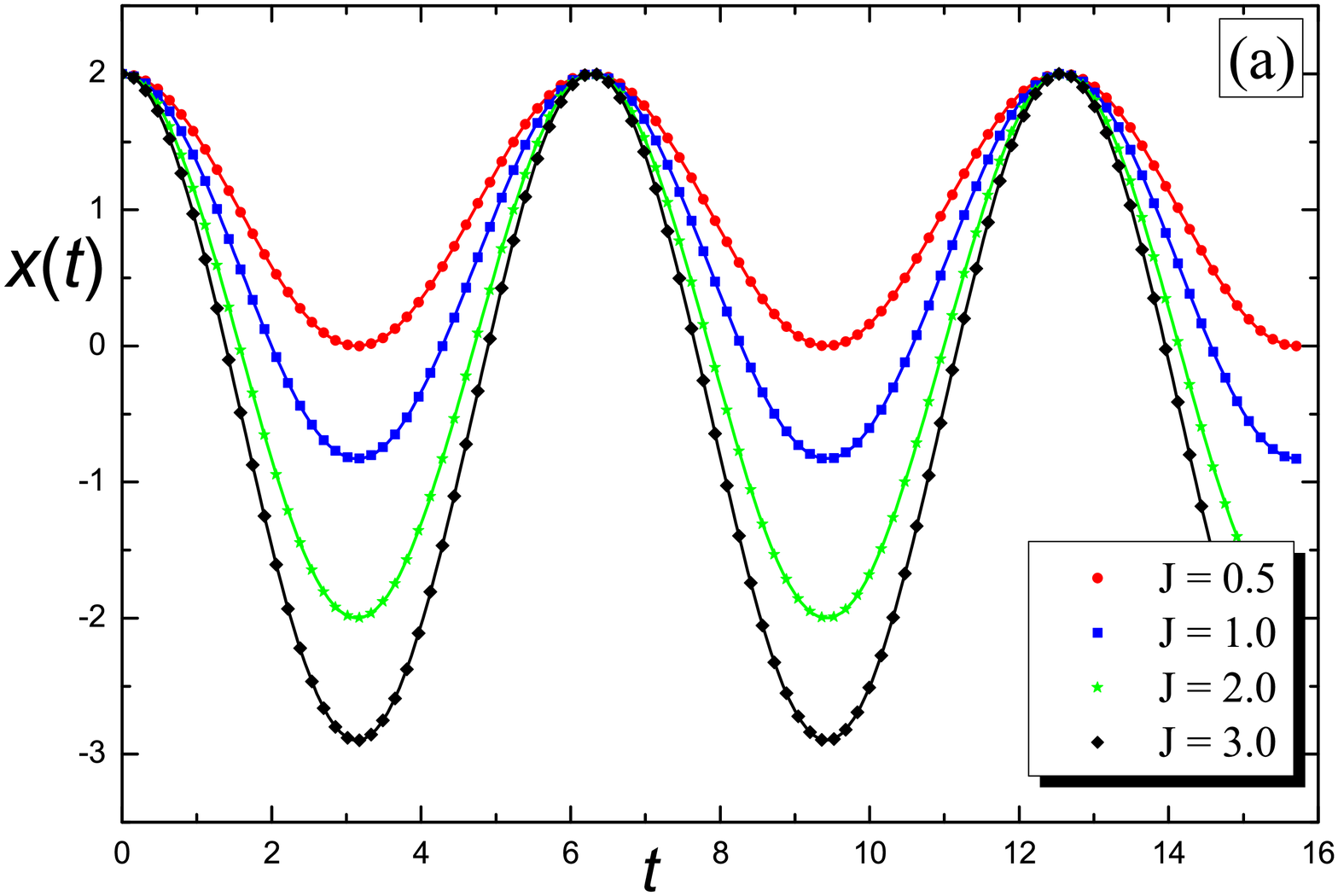} %
\includegraphics[width=7.5cm,height=6.0cm]{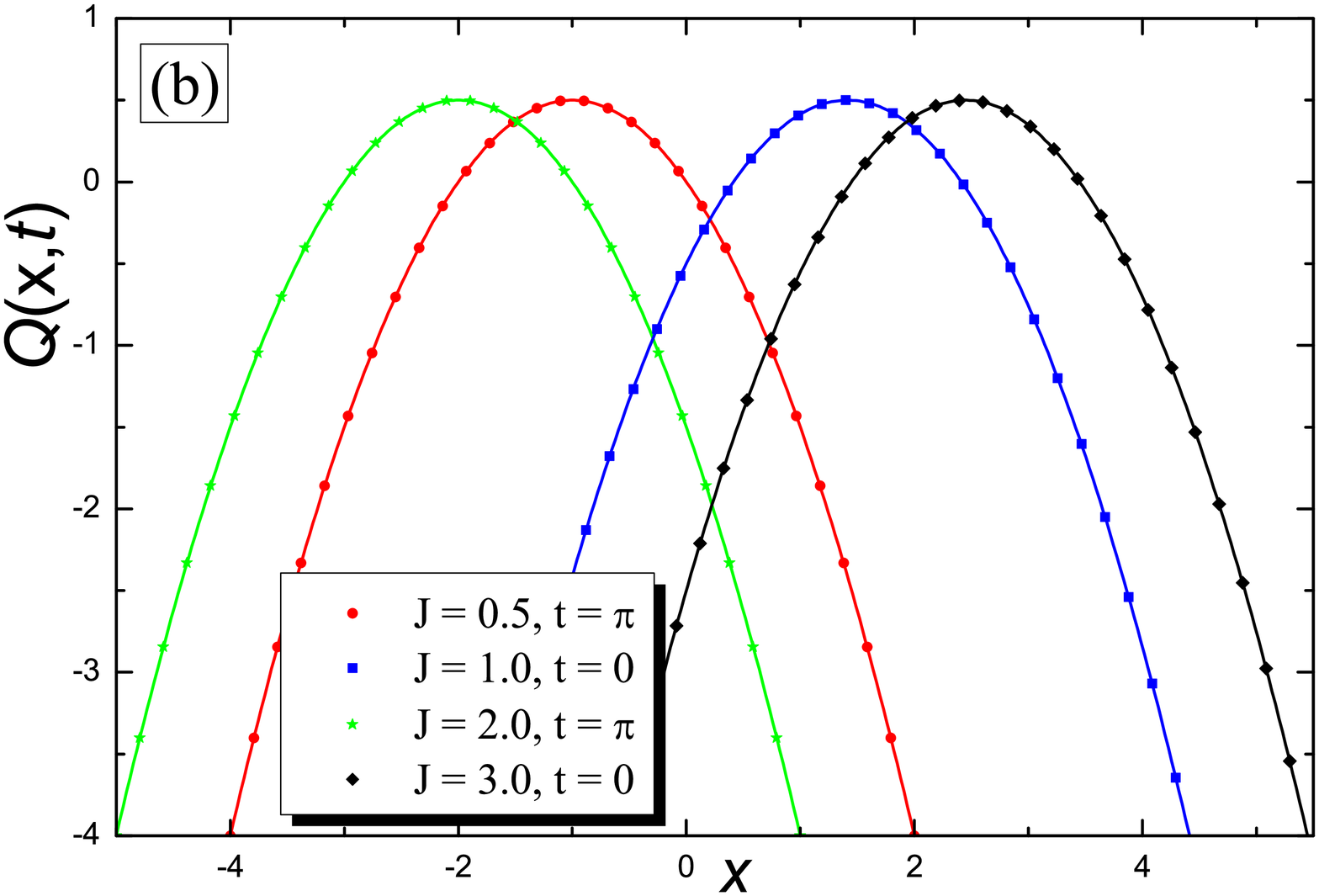}
\caption{(a) Real Bohmian quantum trajectories as functions of time from
Klauder coherent states (scattered) versus classical trajectories
corresponding to (\protect\ref{func}) (solid lines). (b) Quantum potential
from Klauder coherent states (scattered) versus conjectured formula (\protect
\ref{func}) (solid lines). We have taken $x_{0}=2$ and computed the maxima
to $x_{\text{max}}^{0.5}=1$, $x_{\text{max}}^{1}=1.4142$, $x_{\text{max}%
}^{2}=2$, $x_{\text{max}}^{3}=2.4495$.}
\label{F1}
\end{figure}

Due to the fact that $\psi _{J}(x,t)$ involves an infinite sum, it is
complicated to compute analytic expressions for the quantities in (\ref{vQ}%
). However, since we expect a close resemblance to the expressions obtained
from standard coherent states $\psi _{c}(x,t)$, we suggest here that the
corresponding Bohmian trajectories and quantum potential are given by 
\begin{equation}
x(t)=x_{\text{max}}^{J}(\cos \omega t-1)+x_{0}\qquad \text{and\qquad }%
~Q(x,t)=\frac{\hbar \omega }{2}-\frac{1}{2}m\omega ^{2}(x-x_{\text{max}%
}^{J}\cos \omega t)^{2},  \label{func}
\end{equation}%
respectively. Our conjecture is guided by the analogy to the previous case, $%
x_{\text{max}}^{J}$ is taken here to be the centre of the wavepacket, i.e. $%
x_{\text{max}}^{J}=$ $\max \left\vert \psi _{J}(x,0)\right\vert $. In this
case we compute the quantities of interest numerically\footnote{%
If not stated otherwise, we take $\omega =1$, $\hbar =1$ and $m=1$ in all
numerical computations throughout the manuscript.}. We observe stability for 
$\psi _{J}(x,t)$ computed from (\ref{GK}) up to six digits, when terminating
the sum at $n=150$. Our results for the trajectories and quantum potential
are depicted in figure \ref{F1}(a) and \ref{F1}(b), respectively.

We observe perfect agreement between the numerical computation of $x(t)$
from solving the first equation in (\ref{real}) using the expression (\ref%
{GK}) for the Klauder coherent states for various values of $J$ and the
conjectured analytical expression (\ref{func}) for $x(t)$ in which we only
compute the value for $x_{\text{max}}^{J}$ numerically. We find a similar
agreement for the computation of the quantum potential $Q(x,t)$, either
numerically using the expression (\ref{GK}) in the second equation in (\ref%
{real}) or from the conjectured analytical expression in (\ref{func}). We
also find agreement between the two computations solving either directly the
first equation in (\ref{real}) for $x(t)$ or employing the effective
potential $V_{\text{eff}}$ to solve $m\ddot{x}=-\partial V_{\text{eff}%
}/\partial x$ instead. What remains is the interesting challenge to compute
the infinite sums together with the subsequent expressions explicitly in an
analytical manner.

\subsection{Complex case}

As in the previous subsection we start again with stationary states $\psi
_{n}(x,t)=\phi _{n}(x)e^{-iE_{n}t/\hbar }$ as basic input into our
computation as outlined in subsection 2.2. A fundamental difference to the
real case is that now we do not obtain a universal answer for all models.
From (\ref{complex}) we compute%
\begin{eqnarray}
\tilde{v}_{0}(x,t) &=&i\omega x,\qquad \qquad ~~\quad ~~~\tilde{Q}_{0}(x,t)=%
\frac{\hbar \omega }{2},  \label{vq1} \\
\tilde{v}_{1}(x,t) &=&i\omega x-\frac{i\hbar }{mx},~~~~~~\quad ~~\tilde{Q}%
_{1}(x,t)=\frac{\hbar \omega }{2}+\frac{\hbar ^{2}}{2mx^{2}}.  \label{vq2}
\end{eqnarray}%
As discussed in \cite{MVJohn}, for $n=1,2$ the explicit analytical solutions
may be found in these cases. By direct integration of the first equations in
(\ref{vq1}), (\ref{vq2}) or from $m\ddot{x}=-\partial V_{\text{eff}%
}/\partial x$ we compute 
\begin{equation}
x_{0}(t)=x_{0}e^{i\omega t},\quad \text{and\quad }x_{1}(t)=\pm \sqrt{\frac{%
\hbar }{m\omega }+e^{2it\omega }\left( x_{0}^{2}-\frac{\hbar }{m\omega }%
\right) }.
\end{equation}%
For larger values of $n$ we obtain more complicated equations for the
velocities and quantum potentials, which may be solved numerically for $x(t)$%
, see also \cite{MVJohn,ComplYang2}. For instance, we obtain from (\ref%
{complex})%
\begin{eqnarray}
\tilde{v}_{5}(x,t) &=&ix\omega -\frac{5i\hbar }{mx}+\frac{60i\hbar
^{3}-40i\hbar ^{2}mx^{2}\omega }{15\hbar ^{2}mx-20\hbar m^{2}x^{3}\omega
+4m^{3}x^{5}\omega ^{2}}, \\
\tilde{Q}_{5}(x,t) &=&\frac{\hbar \left( 225\hbar ^{5}+225\hbar
^{4}mx^{2}\omega +200\hbar ^{2}m^{3}x^{6}\omega ^{3}-80\hbar
m^{4}x^{8}\omega ^{4}+16m^{5}x^{10}\omega ^{5}\right) }{2m\left( 15\hbar
^{2}x-20\hbar mx^{3}\omega +4m^{2}x^{5}\omega ^{2}\right) ^{2}}~~
\end{eqnarray}%
The solutions for $x_{1}(t)$ and $x_{5}(t)$ are depicted in figure 2.

\begin{figure}[h]
\centering   \includegraphics[width=7.5cm,height=6.0cm]{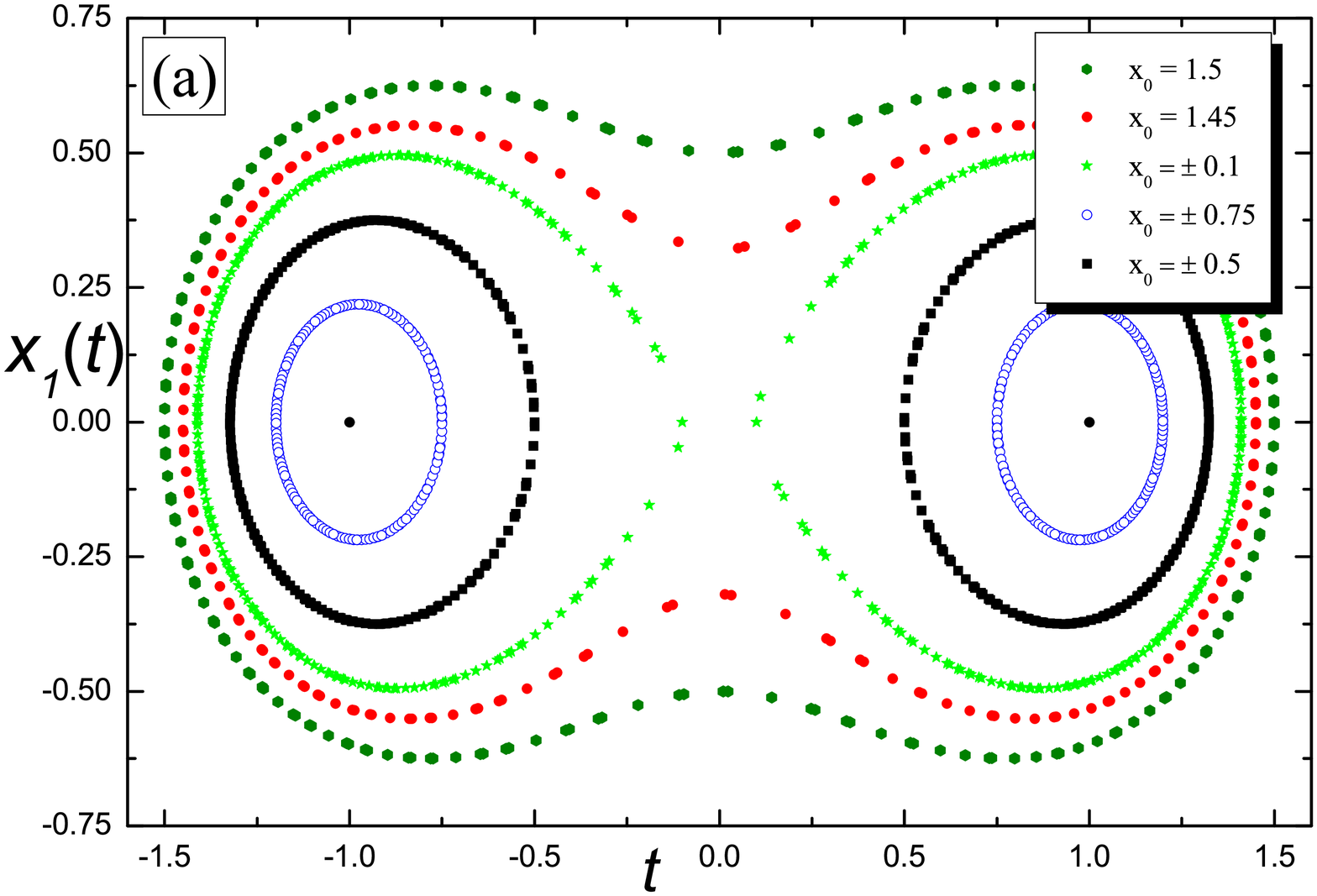} %
\includegraphics[width=7.5cm,height=6.0cm]{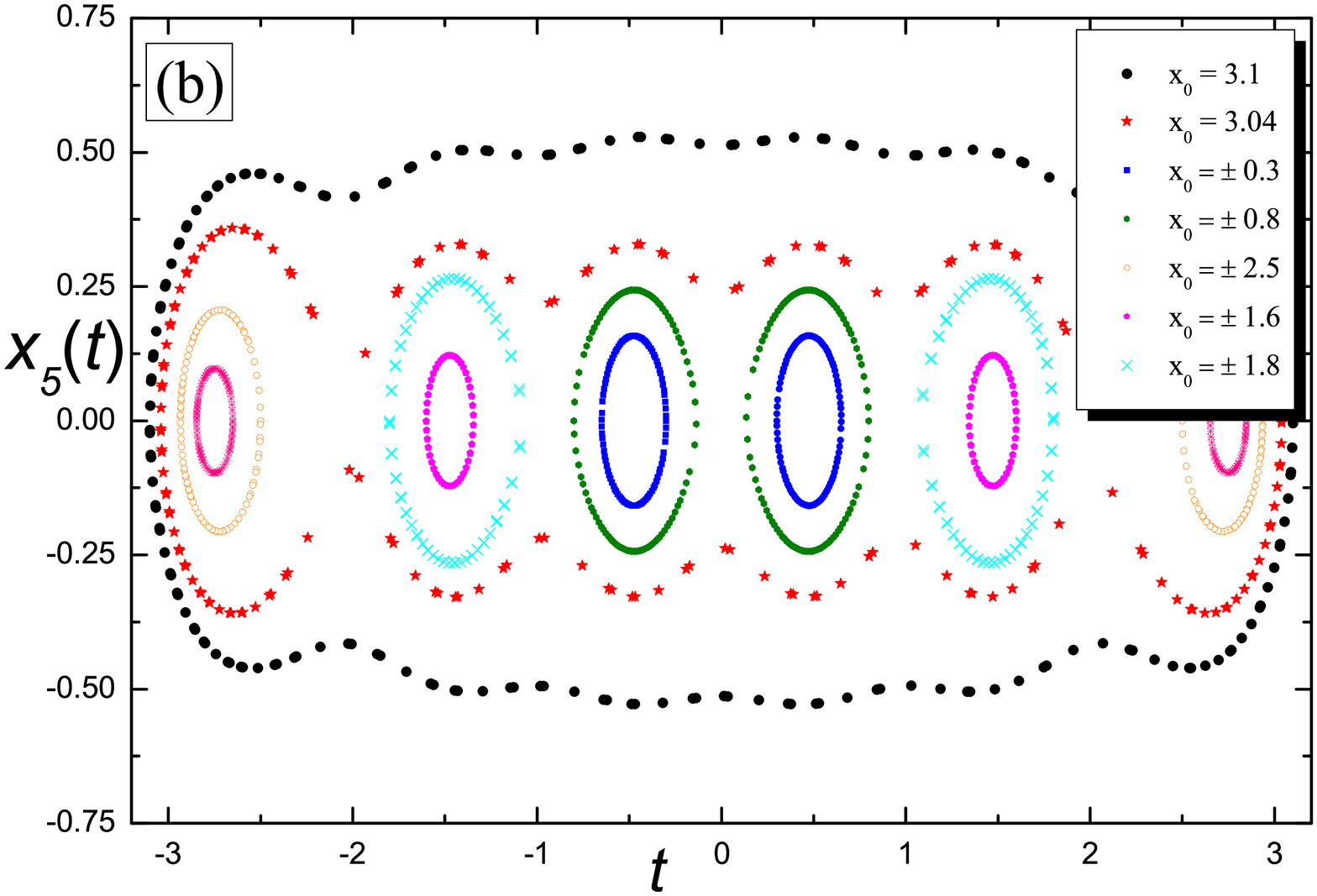}
\caption{Complex Bohmian quantum trajectories as functions of time for
different initial values $x_{0}$ resulting from stationary states $\protect%
\psi _{1}(x,t)$ and $\protect\psi _{5}(x,t)$ in panel (a) and (b),
repectively.}
\label{F2}
\end{figure}

In both cases we observe that the fixed points, at $\pm 1$ for $\tilde{v}%
_{1}(x,t)$ and at $\pm 0.476251$, $\pm 1.47524$, $\pm 2.75624$ for $\tilde{v}%
_{5}(x,t)$, are centres surrounded by closed limit cycles. For large enough
initial values we also observe bounded motion surrounding all fixed points.

Next we use once more the Gaussian wavepackets (\ref{fc}) as input to
evaluate the velocity and the quantum potential from (\ref{complex})%
\begin{equation}
\tilde{v}(x,t)=-\omega (x_{i}+a\sin \omega t)+i\omega (x_{r}-a\cos \omega
t),\qquad \text{and\qquad }~\tilde{Q}(x,t)=\frac{\hbar \omega }{2},
\label{vv}
\end{equation}%
The value for the constant quantum potential was also found in \cite{ChouW}.
Solving now the equation of motion with $\tilde{v}(t)$ for $x_{r}$ and $%
x_{i} $ we obtain a complex trajectory%
\begin{equation}
x(t)=\left( \frac{a}{2}+c_{1}\right) \cos \omega t-c_{2}\sin \omega t+i\left[
c_{2}\cos \omega t+\left( c_{1}-\frac{a}{2}\right) \sin \omega t\right] ,
\label{x1}
\end{equation}%
with integration constants $c_{1}$ and $c_{2}$. We compare this with the
classical result computed from the complex effective Hamiltonian%
\begin{equation}
\mathcal{H}_{\text{eff}}=\frac{1}{2m}(p_{r}^{2}-p_{i}^{2})+\frac{m\omega ^{2}%
}{2}(x_{r}^{2}-x_{i}^{2})+i\left( \frac{1}{m}p_{r}p_{i}+m\omega
^{2}x_{r}x_{i}\right) +\frac{\hbar \omega }{2}.  \label{Heff}
\end{equation}%
We may think of this Hamiltonian as being $\mathcal{PT}$-symmetric, where
the symmetry is induced by the complexification and realized as $\mathcal{PT}
$: $x_{r}\rightarrow -x_{r}$, $x_{i}\rightarrow x_{i}$, $p_{r}\rightarrow
p_{r}$, $p_{i}\rightarrow -p_{i}$, $i\rightarrow -i$. The equations of
motion are then computed according to (\ref{H1}) and (\ref{H2}) to%
\begin{equation}
\dot{x}_{r}=\frac{p_{r}}{m},\quad \dot{x}_{i}=\frac{p_{i}}{m},\quad \dot{p}%
_{r}=-m\omega ^{2}x_{r},\quad \text{and\quad }\dot{p}_{i}=-m\omega ^{2}x_{i}.
\end{equation}%
As these equations decouple, they are easily solved. We find%
\begin{equation}
x(t)=x_{r}(0)\cos \omega t+\frac{p_{r}(0)}{m\omega }\sin \omega t+i\left[
x_{i}(0)\cos \omega t+\frac{p_{i}(0)}{m\omega }\sin \omega t\right] .
\label{x2}
\end{equation}

\begin{figure}[h]
\centering   \includegraphics[width=7.5cm,height=6.0cm]{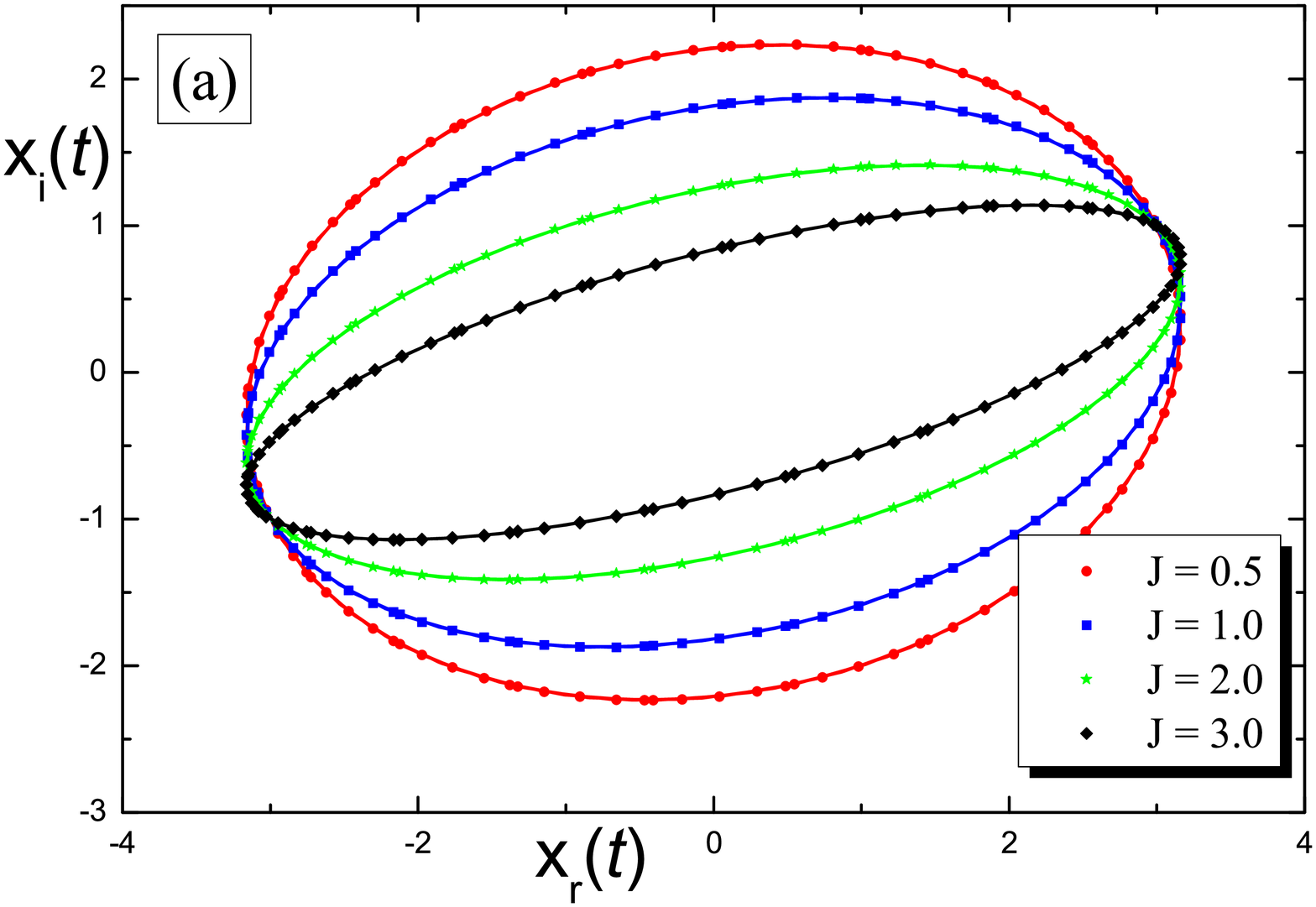} %
\includegraphics[width=7.5cm,height=6.0cm]{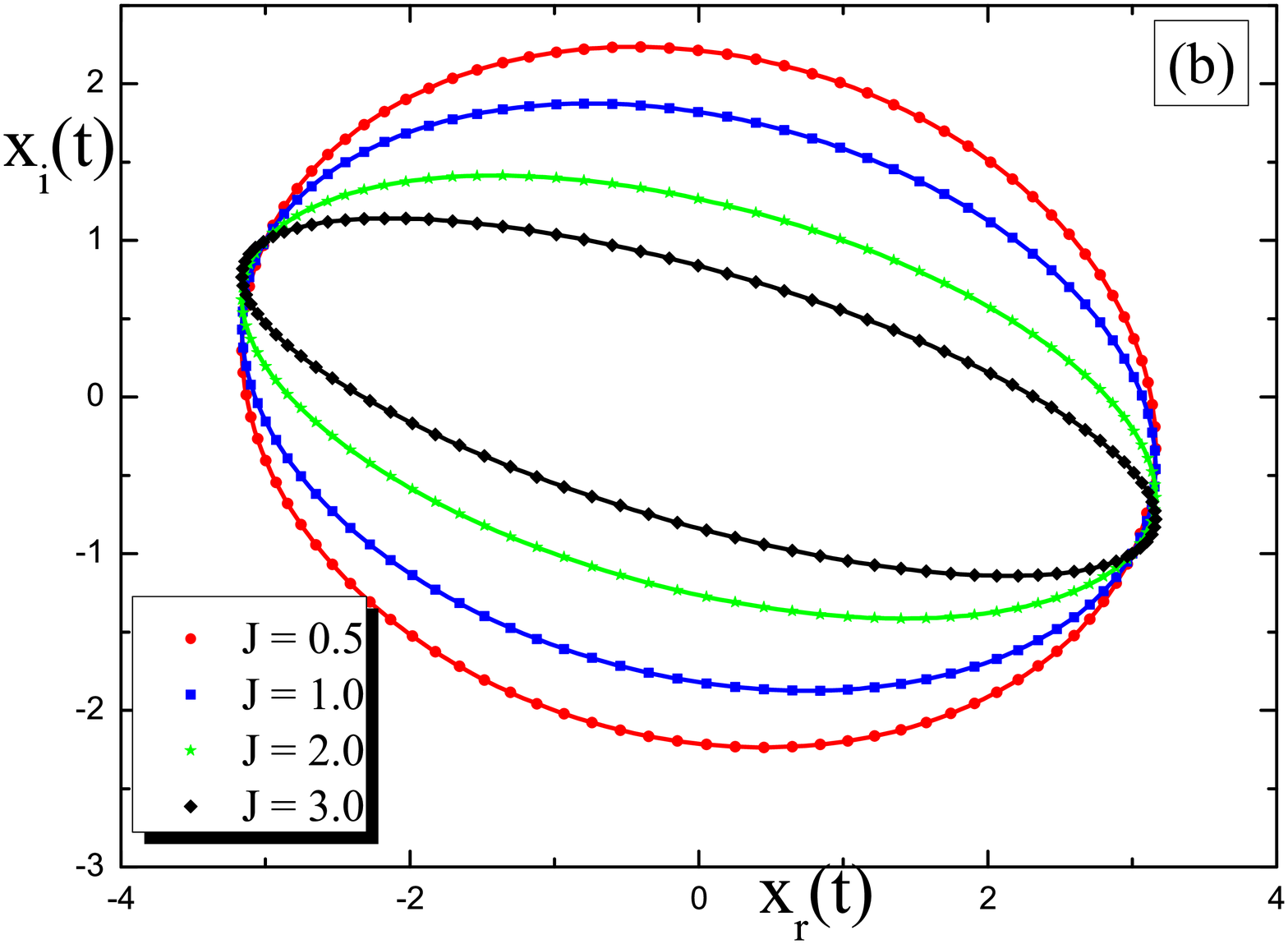}
\caption{Complex Bohmian trajectories resulting from Klauder coherent states
(scattered) compared to the purely classical computation (\protect\ref{x2})
(solid) for different values of $J$, with initial value (a) $x_{0}=3+i$ and
(b) $x_{0}=3-i$ with maximal values $x_{\text{max}}^{0.5}=1$, $x_{\text{max}%
}^{1}=1.4142$, $x_{\text{max}}^{2}=2$, $x_{\text{max}}^{3}=2.4495$.}
\label{F3}
\end{figure}

In order to compare this now with the outcome from taking general Klauder
coherent states (\ref{GK}) in the evaluation of $\tilde{v}(t)$ and $\tilde{Q}%
(x,t)$ and the corresponding trajectories we require the four initial
values. In contrast, solving the first order differential equation for the
velocity (\ref{complex}) we only require the two initial values for the
complex position. To compute the initial values for the momentum we can take
again the results for the Gaussian wavepackets as a guide and compare (\ref%
{x1}) and (\ref{x2}). The compatibility between the two then requires%
\begin{equation}
x_{r}(0)=\frac{p_{i}(0)}{m\omega }+x_{\text{max}}^{J},\qquad \text{and\qquad 
}x_{i}(0)=-\frac{p_{r}(0)}{m\omega },  \label{con}
\end{equation}%
where we have replaced $a$ by $x_{\text{max}}^{J}$. We can now either simply
solve this for the initial values for the momentum (\ref{con}) or
alternatively use directly the same initial values obtained from the
solution of (\ref{complex}). Comparing the direct parametric plot of (\ref%
{x2}) for the stated initial conditions with the numerical computation of
the complex Bohmian trajectories resulting from Klauder coherent states, we
find perfect agreement as depicted in figure \ref{F3}.

Thus under these constraints for the initial conditions the trajectories
resulting from a classical analysis of the effective Hamiltonian (\ref{Heff}%
) and the integration of the complex Bohmian trajectories resulting from
Klauder coherent states are identical. Notice that the quantum nature of $%
\mathcal{H}_{\text{eff}}$ is only visible in form of the overall constant $%
\hbar \omega /2$, which does, however, not play any role in the computation
of the equations of motion.

\section{The P\"{o}schl-Teller potential}

Next we discuss the Bohmian trajectories associated with the P\"{o}%
schl-Teller Hamiltonian \cite{Poschl:1933zz} of the form 
\begin{equation}
\mathcal{H}_{\text{PT}}=\frac{p^{2}}{2m}+\frac{V_{0}}{2}\left[ \frac{\lambda
(\lambda -1)}{\cos ^{2}(x/2a)}+\frac{\kappa (\kappa -1)}{\sin ^{2}(x/2a)}%
\right] -\frac{V_{0}}{2}(\lambda +\kappa )^{2}~~~\text{for }0\leq x\leq a\pi
,  \label{HPT}
\end{equation}%
with $V_{0}=\hbar ^{2}/(4ma^{2})$. This model has been widely discussed in
the mathematical physics literature, e.g. \cite{Kleinert,Klauder2}, since it
has the virtue of being exactly solvable, classically as well as quantum
mechanically. For a given energy $E$ a classical solution is known to be%
\begin{equation}
x(t)=a\arccos \left[ \frac{\alpha -\beta }{2}+\sqrt{\gamma }\cos \left( 
\sqrt{\frac{2E}{m}}\frac{t}{a}\right) \right] ,  \label{PTcl}
\end{equation}%
with $\alpha =\lambda (\lambda -1)V_{0}/E$, $\beta =\kappa (\kappa
-1)V_{0}/E $ and $\gamma =\alpha ^{2}/4+\beta ^{2}/4-\alpha \beta /2-\alpha
-\beta +1$. The time dependent Schr\"{o}dinger equation is solved by
discrete eigenfunctions 
\begin{equation}
\psi _{n}(x,t)=\frac{1}{\sqrt{N_{n}}}\cos ^{\lambda }\left( \frac{x}{2a}%
\right) \sin ^{\kappa }\left( \frac{x}{2a}\right) ~_{2}F_{1}\left[
-n,n+\kappa +\lambda ;k+\frac{1}{2};\sin ^{2}\left( \frac{x}{2a}\right) %
\right] e^{-iE_{n}t/\hbar }  \label{PTST}
\end{equation}%
$\allowbreak $with $_{2}F_{1}$ denoting the Gauss hypergeometric function.
The corresponding energy eigenvalues and the normalization factor are given
by 
\begin{equation}
E_{n}=\frac{\hbar ^{2}}{2ma^{2}}n(n+\kappa +\lambda ),\text{\quad }%
N_{n}=a2^{n}n!\frac{\Gamma (\kappa +1/2)\Gamma (n+\lambda +1/2)}{\Gamma
(2n+1+\lambda +\kappa )}\dprod\limits_{l=1}^{n}\frac{n-1+l+\kappa +\lambda }{%
2l-1+2\kappa },
\end{equation}%
respectively. We will use these solutions in what follows.

\subsection{Real case}

As in the previous case we start with the construction of the trajectories
from stationary states (\ref{PTST}). Once again for the real case the
computation is unspectacular in this case since the velocity computed from (%
\ref{real}) is $v(t)=0$ and the corresponding quantum potential results
again simply to $Q(x)=E_{n}-V_{\text{PT}}(x)$, such that classical
trajectories correspond to a motion in a constant effective potential $V_{%
\text{eff}}(x,t)=E_{n}$.

More interesting, and qualitatively very close to the classical behaviour,
are the trajectories resulting from the Klauder coherent states given by the
general expression (\ref{GK}). In this case the probability distribution is
computed with $e_{n}=n(n+\kappa +\lambda )$ to $\rho _{n}=n!(n+\kappa
+\lambda )_{n}$, where $(x)_{n}:=\Gamma (x+n)/\Gamma (x)$ denotes the
Pochhammer symbol. With these expressions the normalization constant results
to a confluent hypergeometric function $\mathcal{N}^{2}(J)=~_{0}F_{1}\left(
1+\kappa +\lambda ;J\right) $, from which we compute the Mandel parameter (%
\ref{Mandel}) to 
\begin{equation}
Q(J,\kappa +\lambda )=\frac{J}{2+\kappa +\lambda }~\frac{_{0}F_{1}\left(
3+\kappa +\lambda ;J\right) }{_{0}F_{1}\left( 2+\kappa +\lambda ;J\right) }-%
\frac{J}{1+\kappa +\lambda }~\frac{_{0}F_{1}\left( 2+\kappa +\lambda
;J\right) }{_{0}F_{1}\left( 1+\kappa +\lambda ;J\right) }~.
\end{equation}%
Using the relation between the confluent hypergeometric function and the
modified Bessel function this is easily converted into the expression found
in \cite{Klauder2}. We agree with the finding therein that $Q$ is always
negative, but disagree with the statement that $Q$ tends to zero for large $%
J $ for fixed $\kappa $, $\lambda $. Instead we argue that for fixed
coupling constants the Mandel parameter $Q$ is a monotonically decreasing
function of $J$ with $Q(0,\kappa +\lambda )=0$. Assuming that the coherent
states closely resemble a classical behaviour, we conjecture here in analogy
to the classical solution (\ref{PTcl}) that the quantum trajectories acquire
the general form 
\begin{equation}
x(t)=a\arccos \left[ \frac{X_{+}}{2}+\frac{X_{-}}{2}\cos \left( 2\pi \frac{t%
}{T}\right) \right] ,  \label{PTx}
\end{equation}%
with $X_{\pm }=\cos (x_{0}/a)\pm \cos (x_{m}/a)$, $T$ denoting the period
and $x_{m}=x(T/2)=\max [x(t)]$. Our conjecture is based on an extrapolation
of the analysis of the relations between $\alpha $ and $\beta $ and
functions of $x(0)$ and $x(T/2)$. The effective potential computed from (\ref%
{PTx}) is then of P\"{o}schl-Teller type 
\begin{equation}
V_{\text{eff}}=\frac{2ma^{2}\pi ^{2}}{T^{2}}\left[ \frac{\cos
^{2}(x_{0}/2a)\cos ^{2}(x_{m}/2a)}{\cos ^{2}(x/2a)}+\frac{\sin
^{2}(x_{0}/2a)\sin ^{2}(x_{m}/2a)}{\sin ^{2}(x/2a)}\right] .
\end{equation}

\begin{figure}[h]
\centering   \includegraphics[width=7.5cm,height=6.0cm]{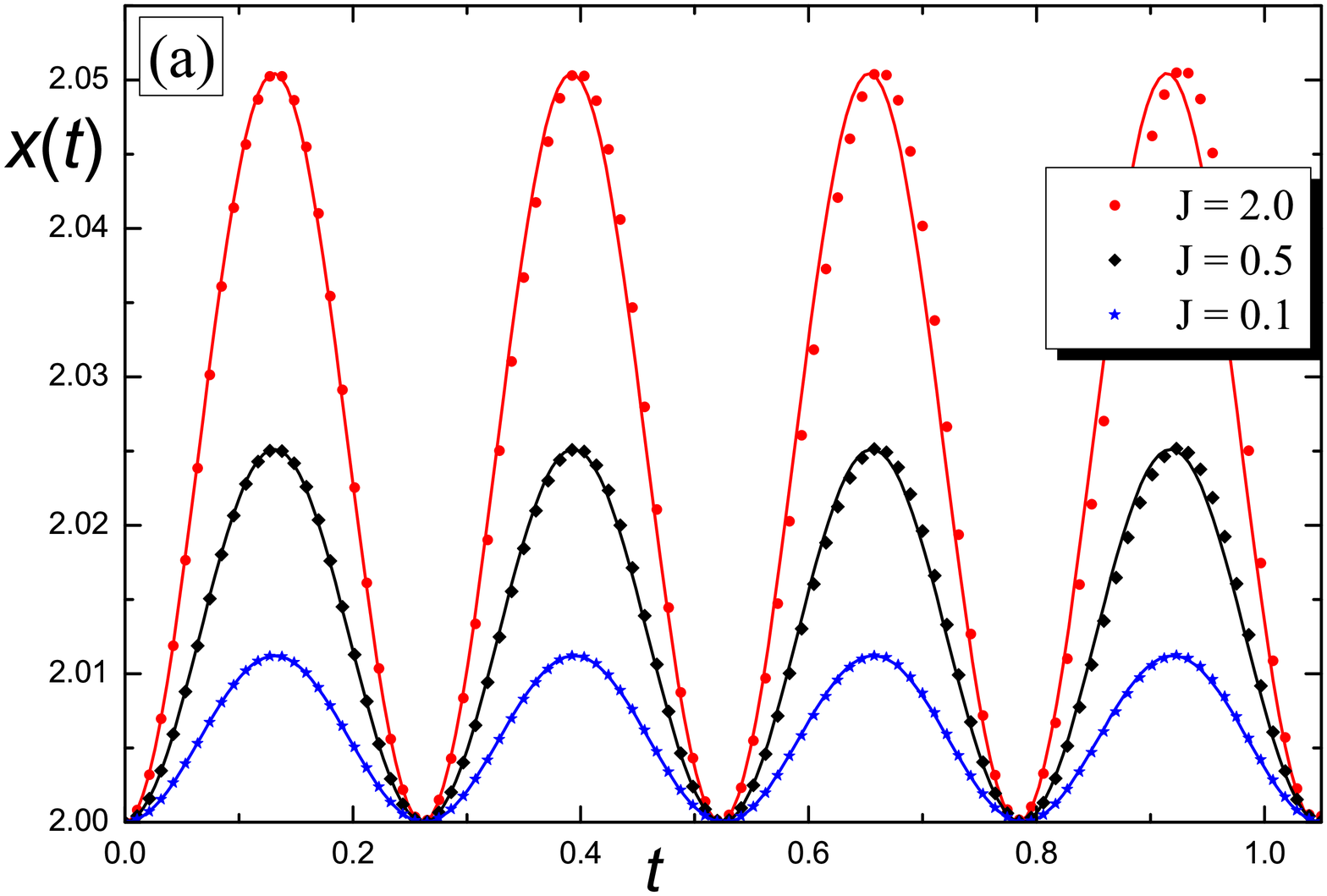} %
\includegraphics[width=7.5cm,height=6.0cm]{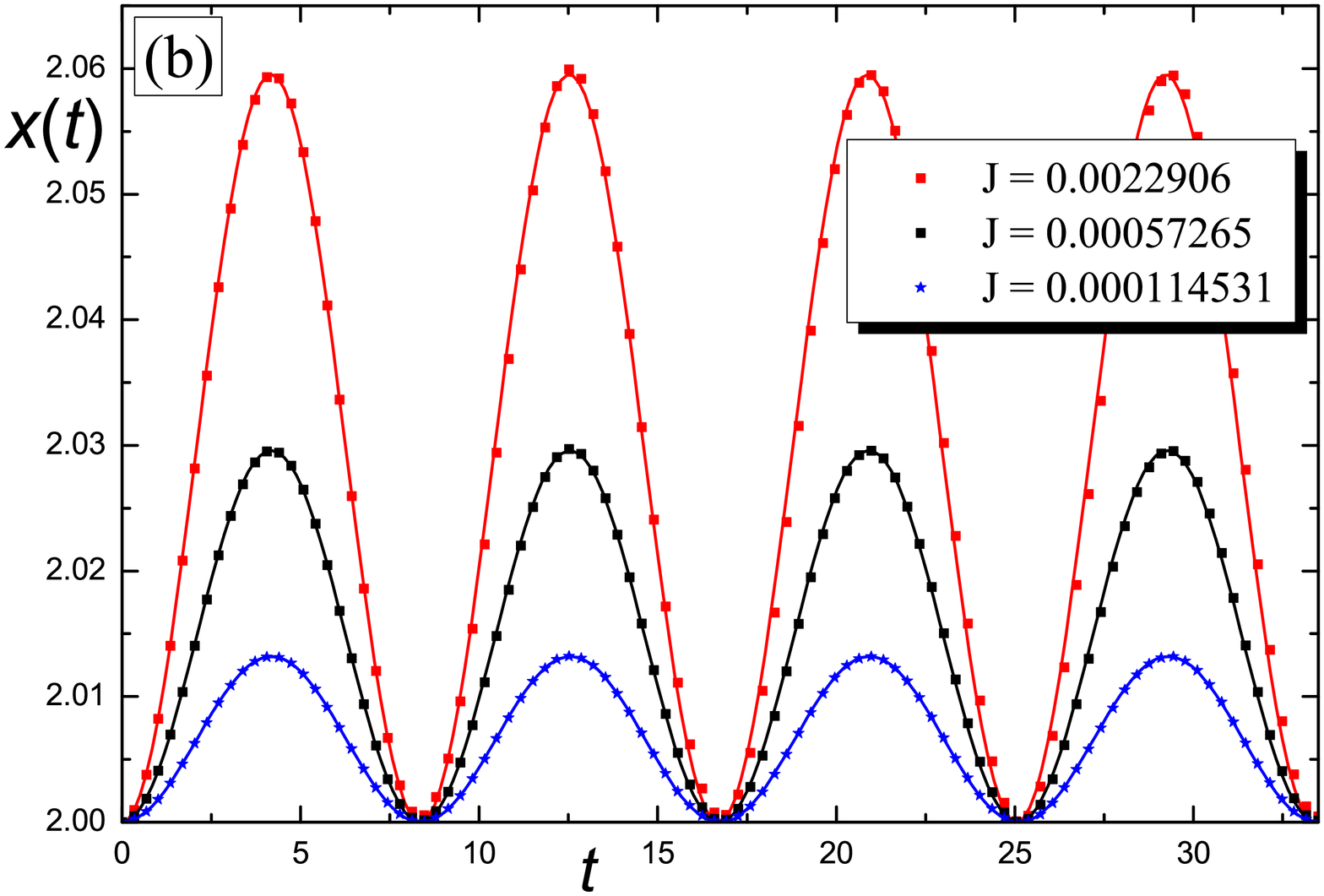} \centering   %
\includegraphics[width=7.5cm,height=6.0cm]{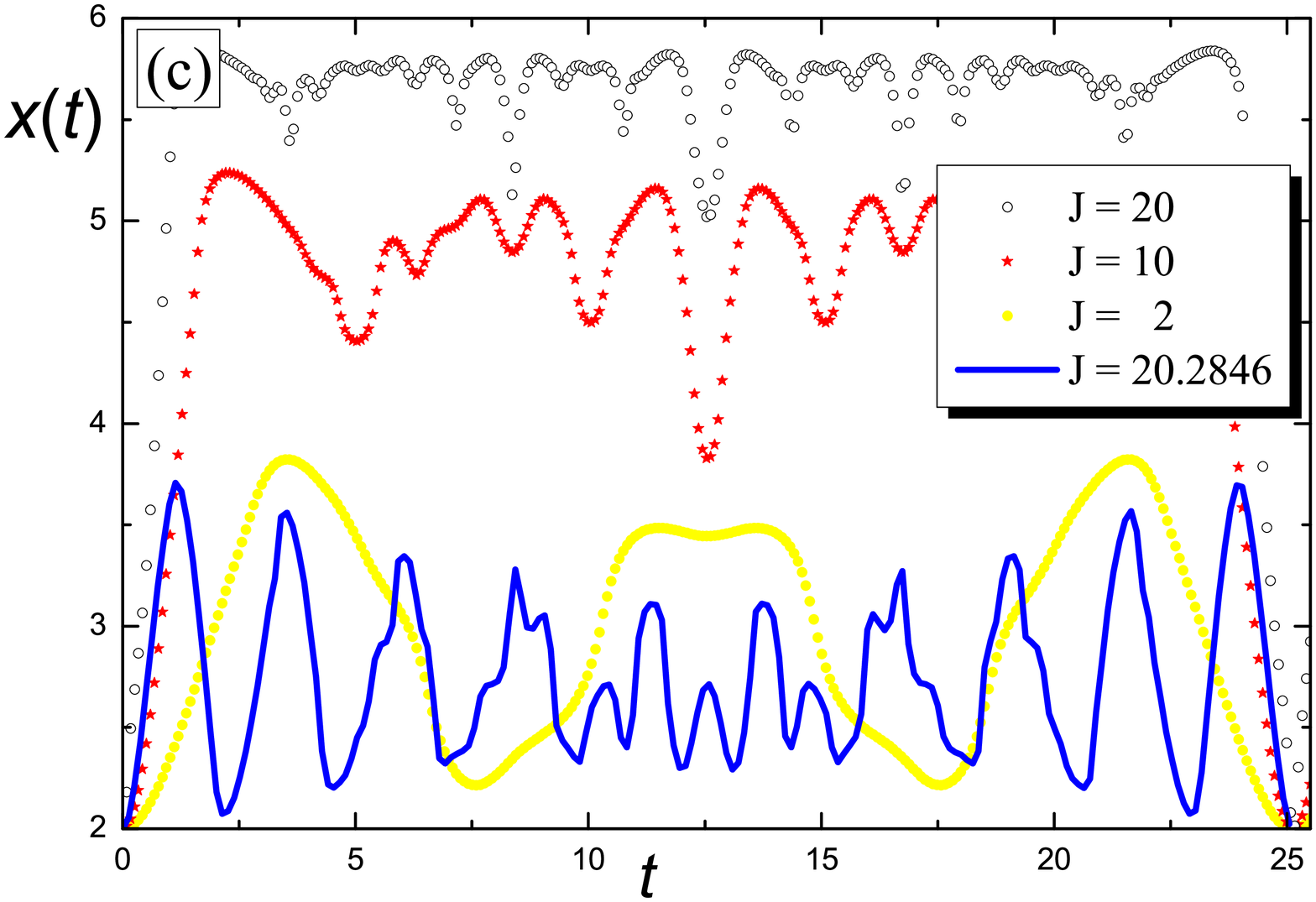} %
\includegraphics[width=7.5cm,height=6.0cm]{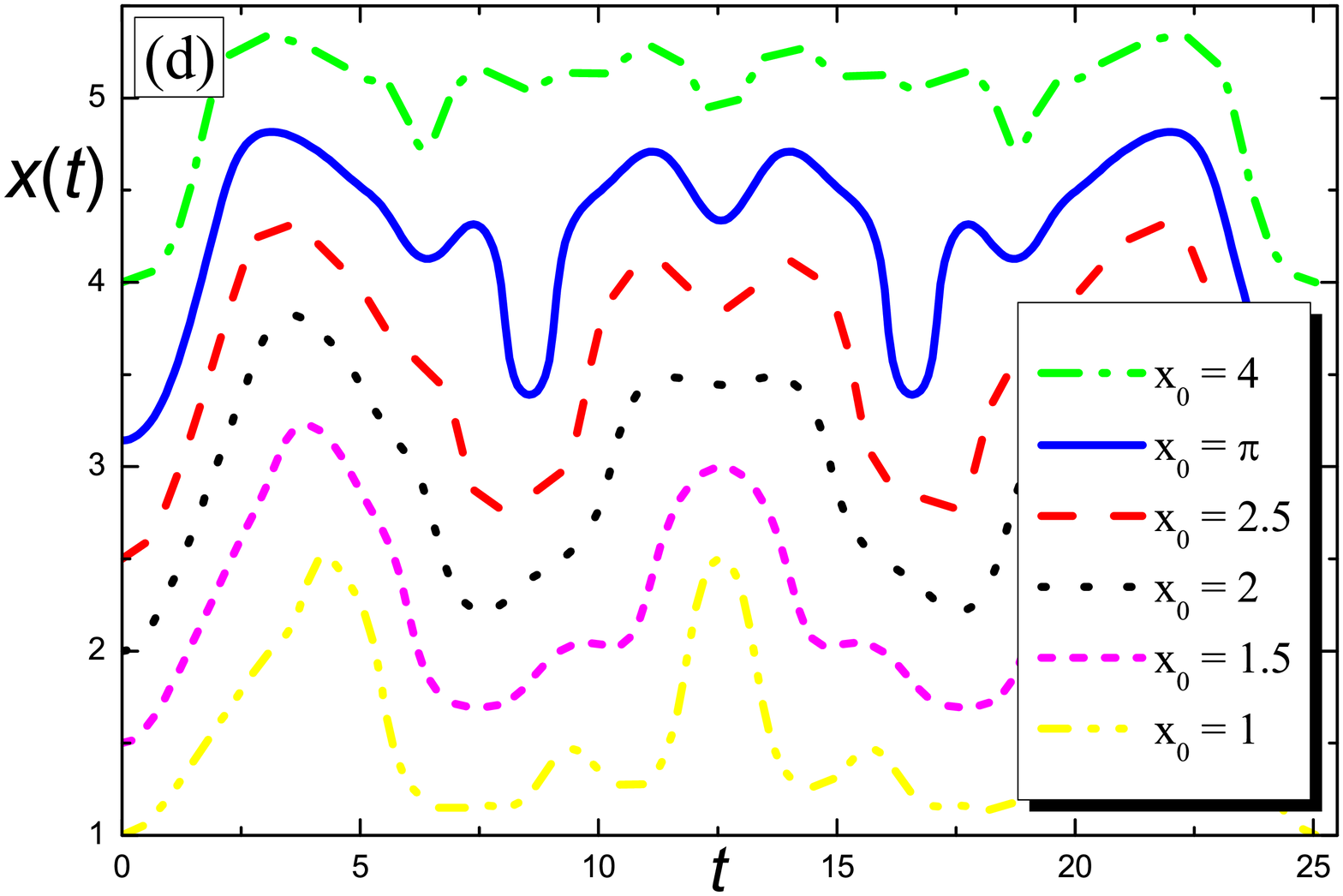}
\caption{ Real Bohmian trajectories as functions of time from Klauder
coherent states (scattered) versus classical trajectories (solid lines)
corresponding to (\protect\ref{PTx}). (a) Quasi-Poissonian distribution with
initial value $x_{0}=2$, coupling constants $\protect\kappa =90$, $\protect%
\lambda =100$, maxima $x_{\text{m}}^{2}=2.0504447$, $x_{\text{m}%
}^{0.5}=2.0251224$, $x_{\text{m}}^{0.1}=2.0112100$ and periods $%
T^{2}=0.2612875$, $T^{0.5}=0.2622200$, $T^{0.1}=0.2627495$ for different
values of $J$. (b) Quasi-Poissonian distribution with initial value $x_{0}=2$%
, coupling constants $\protect\kappa =2$, $\protect\lambda =3$, maxima $x_{%
\text{m}}^{0.0022906}=2.059522$, $x_{\text{m}}^{0.00057265}=2.0295876$, $x_{%
\text{m}}^{0.000114531}=2.0131884$ and periods $T^{0.0022906}=8.34795$, $%
T^{0.00057265}=8.36305$, $T^{0.000114531}=8.37129$ for different values of $%
J $. (c) Sub-Poissonian distribution with initial value $x_{0}=2$, coupling
constants $\protect\kappa =2$, $\protect\lambda =3$ for $J=2,10,20 $
(scattered) and $\protect\kappa =9$, $\protect\lambda =10$ for $J=20.2846$
(solid). (d) Sub-Poissonian distribution for various initial values with
coupling constants $\protect\kappa =2$, $\protect\lambda =3$ for $J=2$.}
\label{F4}
\end{figure}

As in the previous case we will compute the quantum trajectories numerically%
\footnote{%
We take $a=2$ in all numerical computations in this section.}. Our results
from solving (\ref{real}) are depicted in figure \ref{F4}.

Most importantly we observe that the behaviour of the trajectories is
entirely controlled by the values of the Mandel parameter $Q$. Panel (a) and
(b) show trajectories for different values of $J$ with pairwise identical
values of the Mandel parameter, that is $%
Q(2,190)=Q(0.0022906,5)=-0.000054529 $, $%
Q(0.5,190)=Q(0.00057265,5)=-0.000013634$ and $%
Q(0.1,190)=Q(0.000114531,5)=-2.72691\times 10^{-6}$. We notice that the
overall qualitative behaviour is simply rescaled in time. We further observe
a small deviation from the periodicity growing with increasing time. As a
consequence the matching between the quantum trajectories obtained from
solving (\ref{real}) and our conjectured analytical expression (\ref{PTx})
is good for small values of time, but worsens as time increases. The
agreement improves the closer the Mandel parameter approaches the Poissonian
distribution, i.e. $Q=0$. Once the Mandel parameter becomes very negative
the correlation between the classical motion and the Bohmian trajectories is
entirely lost as shown in panel (c) of figure \ref{F4} for $%
Q(2,5)=-0.0425545 $, $Q(10,5)=-0.149523$ and $Q(20,5)=-0.218944$. We also
notice from panel (c) that the qualitative similarity observed for equal
values of the Mandel parameter seen in panels (a) and (b) is lost once the
states do not resemble a classical behaviour. This is seen by comparing \
the yellow dotted line and the solid blue line corresponding to the same
values $Q(2,5)=Q(20.2846,19)=-0.0425545$. Panel (d) shows the sensitivity
with regard to the initial values $x_{0}$. Whereas for the trajectories
resembling the classical motion (\ref{PTx}) this change does not affect the
overall qualitative behaviour, it produces a more significant variation in
the non-classical regime.

\begin{figure}[h]
\centering   \includegraphics[width=7.5cm,height=6.0cm]{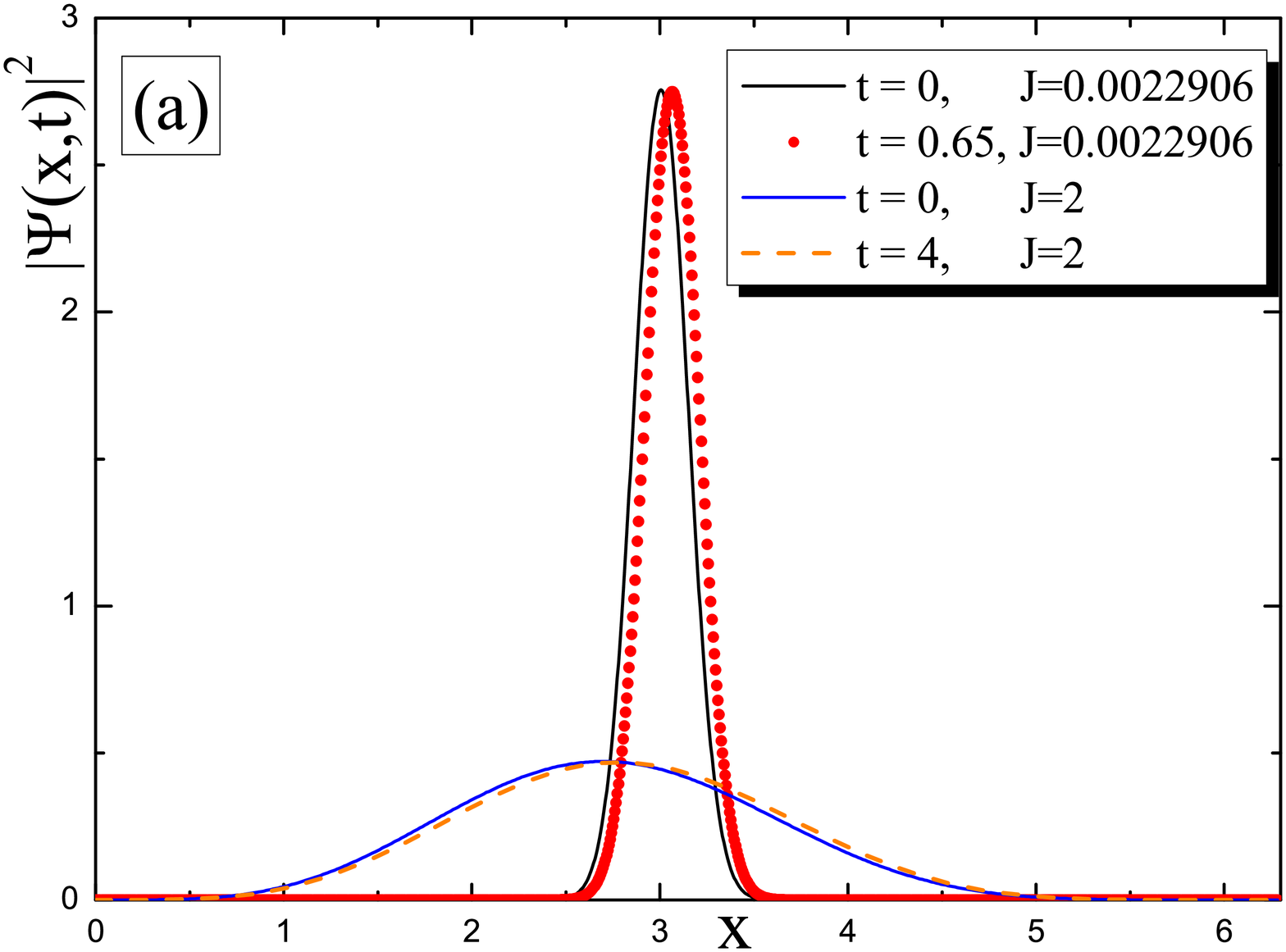} %
\includegraphics[width=7.5cm,height=6.0cm]{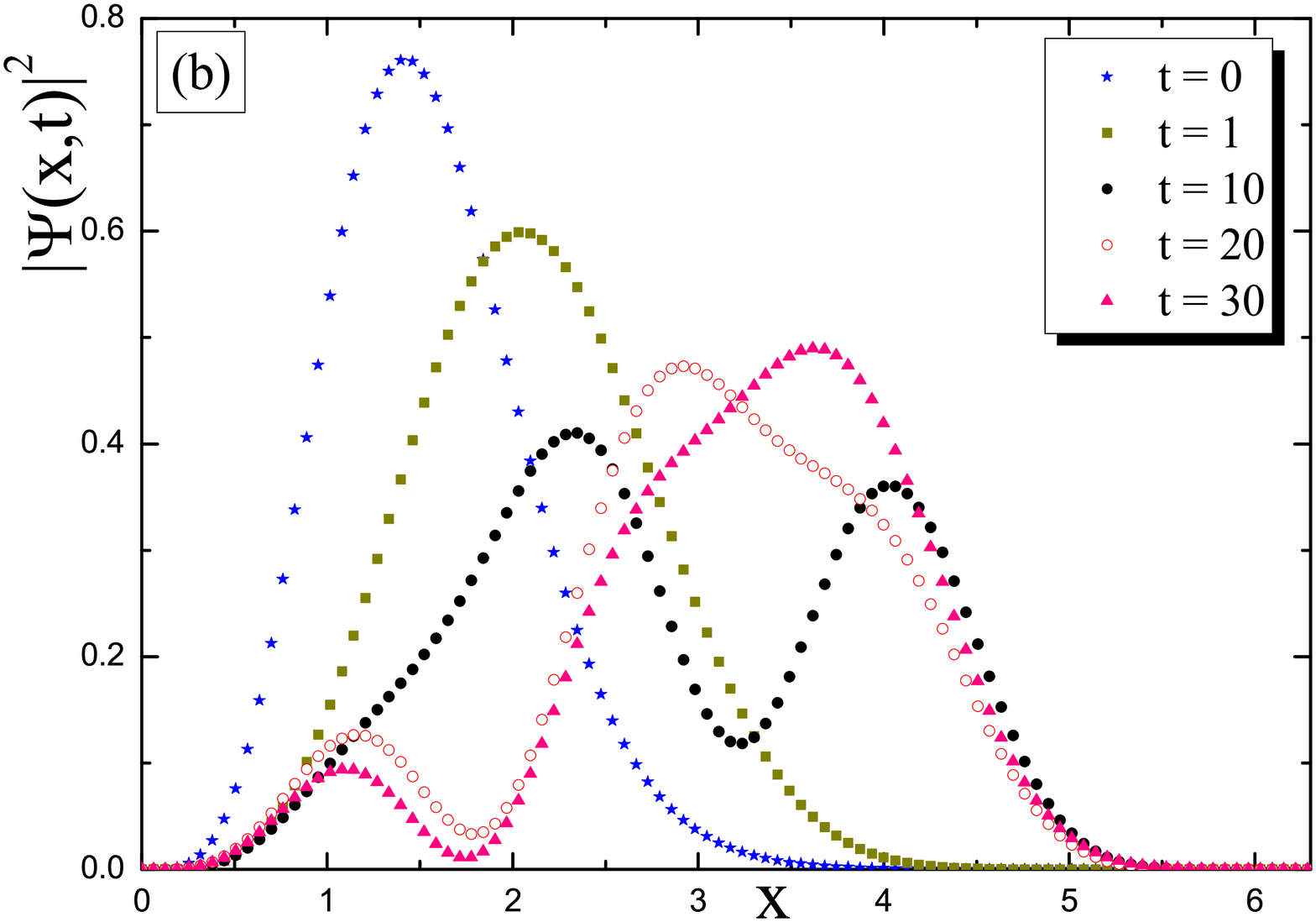}
\caption{(a) Periodic soliton like motion in the quasi-Poissonian regime for 
$\protect\kappa =90$, $\protect\lambda =100$ (thin) and $\protect\kappa =2$, 
$\protect\lambda =3$ (broad) with $Q=-0.000054529$ identical in both cases.
(b) Spreading wave in the sub-Poissonian regime with $Q=-0.0917752$ for $J=5$
and $\protect\kappa =2$, $\protect\lambda =3$. }
\label{F7}
\end{figure}

The explanation for this behaviour is that in the quasi-Poissonian regime
the coherent states evolve as soliton like structures keeping their shape
carrying out a periodic motion in time. In contrast, in the sub-Poissonian
regime the motion is no longer periodic and the initial Gaussian shape of
the wave is dramatically changed under the evolution of time. These features
are demonstrated in figure \ref{F7}.

\begin{figure}[h]
\centering   \includegraphics[width=7.5cm,]{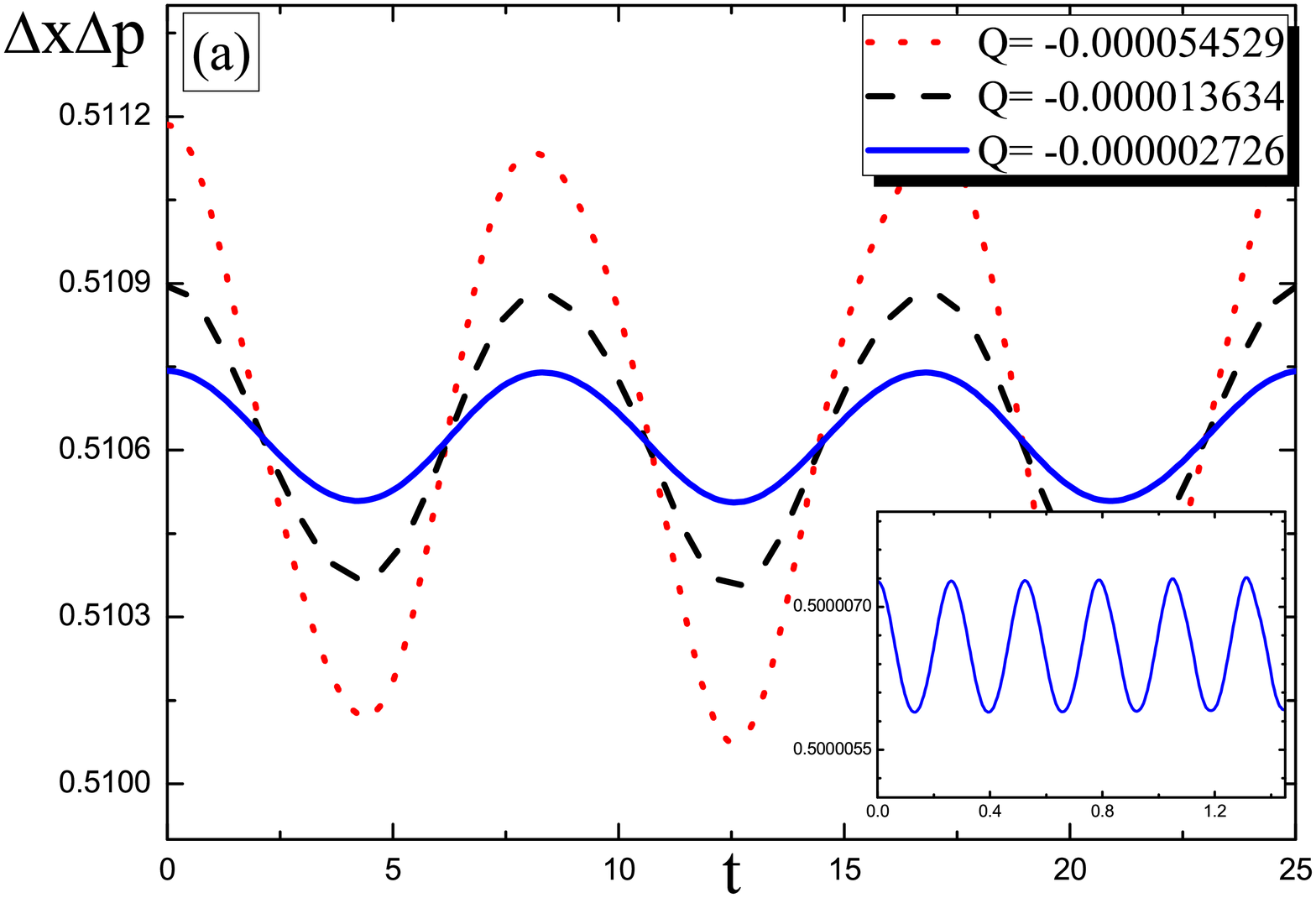} %
\includegraphics[width=7.5cm]{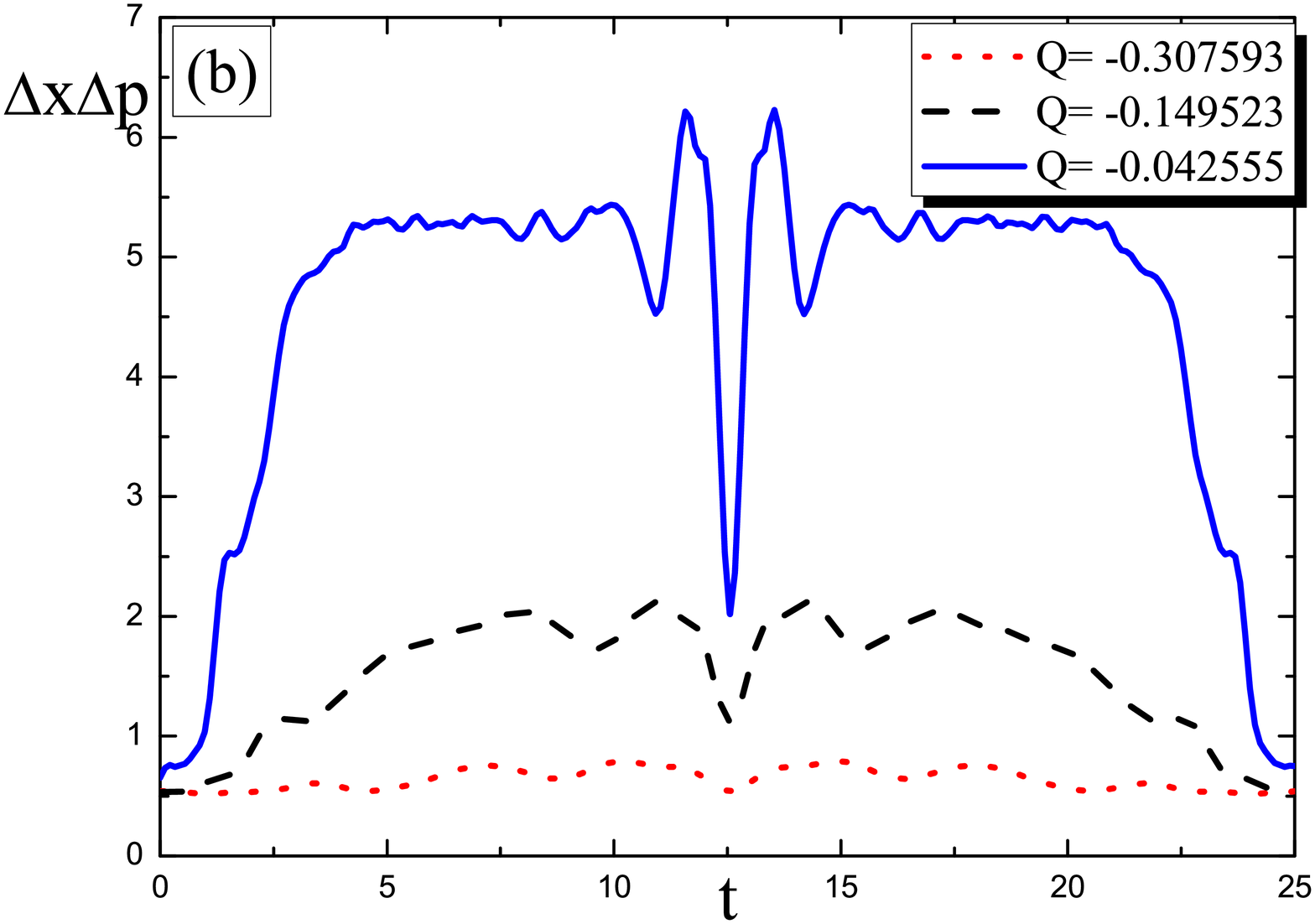}
\caption{Product of the position and momentum uncertainty as functions of
time for different values of the Mandel parameter $Q$. (a) The coupling
constants are $\protect\kappa =2,\protect\lambda =3$, with $J=0.0022906$
(red dotted), $J=0.00057265$ (black dashed), $J=0.000114531$ (blue solid)
and $J=0.1$ for the subpanel with $\protect\kappa =90,\protect\lambda =100$.
(b) The coupling constants are $\protect\kappa =2,\protect\lambda =3$ with $%
J=2$ (red dotted), $J=10$ (black dashed) and $J=50$ (blue solid).}
\label{F8}
\end{figure}

\noindent In figure \ref{F8} we plot the uncertainty relations for
comparison.

In panel (a) we observe that in the quasi-Poissonian regime the saturation
level is almost reached with $\Delta x \Delta p$ being very close to $\hbar
/ 2$, oscillating around $0.5106$ with a deviation of $\pm 0.0006$ and in
the subpanel oscillating around $0.5000065$ with a deviation of $\pm
0.0000008$. This is of course compatible with the very narrow soliton like
structure observed in figure \ref{F7} leading to a classical type of
behaviour. However, in the sub-Poissonian regime the uncertainty becomes
larger, as seen in panel (b), corresponding to a spread out wave behaving
very non-classical.

\subsection{Complex case}

\begin{figure}[ht]
\centering   \includegraphics[width=7.5cm,height=6.0cm]{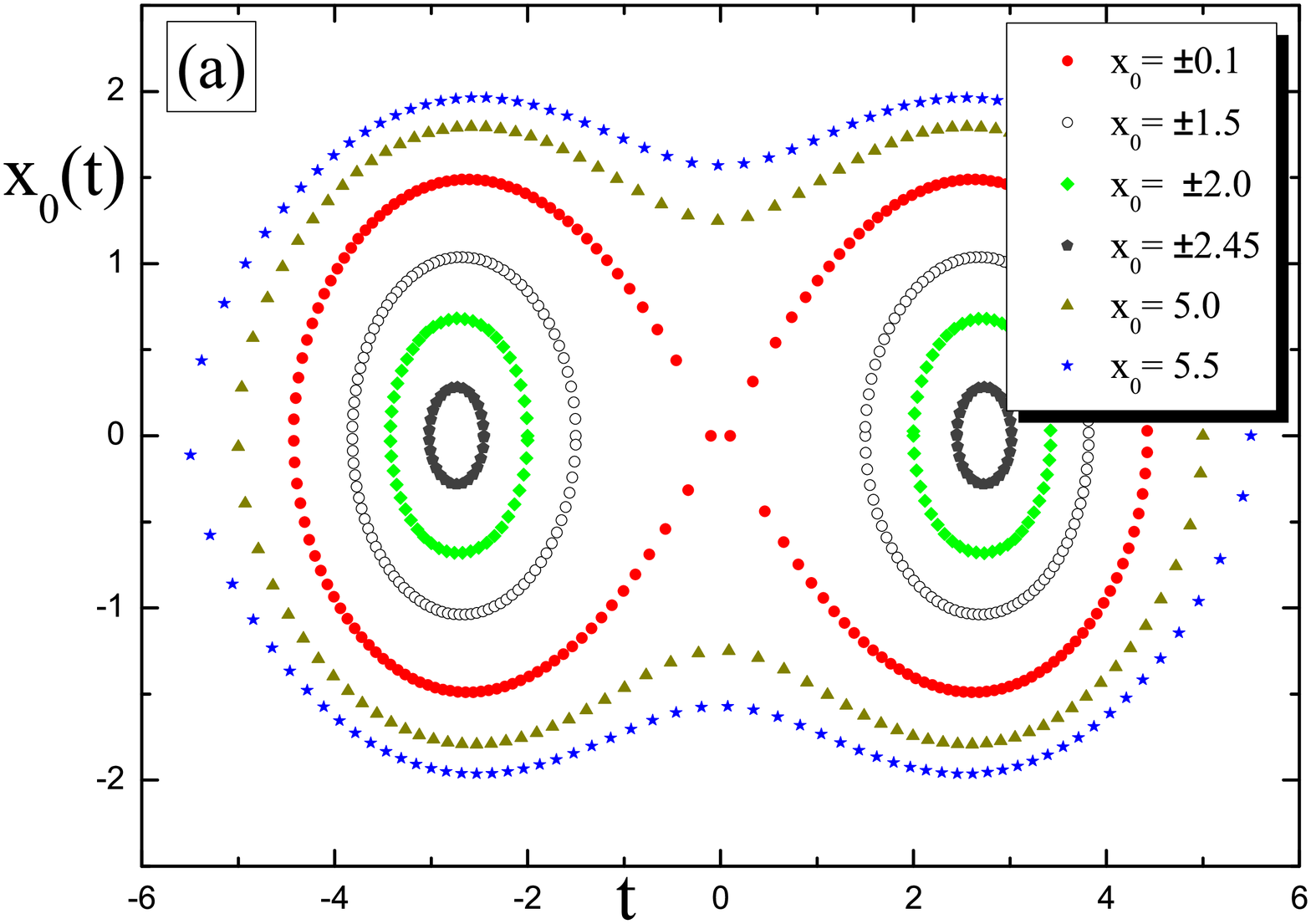} %
\includegraphics[width=7.5cm,height=6.0cm]{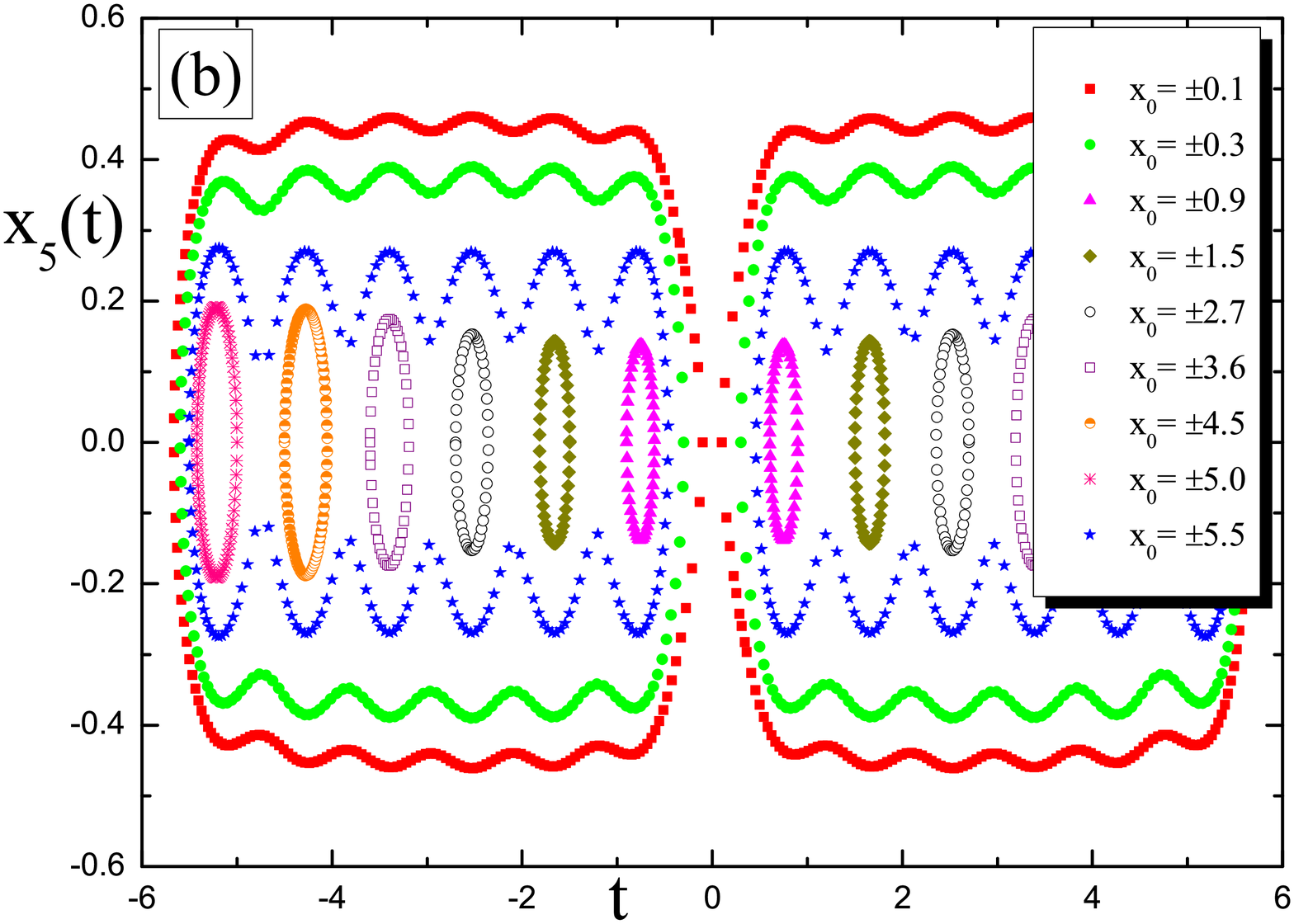}
\caption{Complex Bohmian trajectories as functions of time for different
initial values $x_{0}$ resulting from stationary states $\Psi _{0}(x,t)$ and 
$\Psi _{5}(x,t)$ in panel (a) and (b), respectively.}
\label{F6}
\end{figure}

Let us now consider the complex Bohmian trajectories starting once again
with the construction from stationary states $\psi _{n}(x,t)=\phi
_{n}(x)e^{-iE_{n}t/\hbar }$. For the lowest states we may compute analytical
expressions from (\ref{complex}) for the velocities 
\begin{eqnarray}
\tilde{v}_{0}(x,t) &=&\frac{\hbar \left[ (\kappa +\lambda )\cos \left( \frac{%
x}{a}\right) +\kappa -\lambda \right] }{i2am\sin \left( \frac{x}{a}\right) }%
,~~ \\
\tilde{v}_{1}(x,t) &=&\frac{\hbar \left[ (2\kappa ^{2}+\kappa )\cot \left( 
\frac{x}{2a}\right) +(2\lambda ^{2}+\lambda )\tan \left( \frac{x}{2a}\right)
-(\kappa +\lambda +1)(\kappa +\lambda +2)\sin \left( \frac{x}{a}\right) %
\right] }{i2am\left[ (\kappa +\lambda +1)\cos \left( \frac{x}{a}\right)
+\kappa -\lambda \right] },  \notag
\end{eqnarray}%
and the quantum potentials%
\begin{eqnarray}
\tilde{Q}_{0}(x,t) &=&V_{0}\frac{\left[ (\kappa -\lambda )\cos \left( \frac{x%
}{a}\right) +\kappa +\lambda \right] }{\sin ^{2}\left( \frac{x}{a}\right) },
\label{Q1} \\
\tilde{Q}_{1}(x,t) &=&\frac{V_{0}}{2}\left[ \frac{4(\kappa +\lambda
+1)\left( (\kappa -\lambda )\cos \left( \frac{x}{a}\right) +\kappa +\lambda
+1\right) }{\left[ (\kappa +\lambda +1)\cos \left( \frac{x}{a}\right)
+\kappa -\lambda \right] ^{2}}+\frac{\kappa }{\sin ^{2}\left( \frac{x}{2a}%
\right) }+\frac{\lambda }{\cos ^{2}\left( \frac{x}{2a}\right) }\right] . 
\notag  \label{Q2}
\end{eqnarray}%
We note that the quantum potential $\tilde{Q}_{1}(x,t)$ resembles a P\"{o}%
schl-Teller potential apart from its first term. For $n=0$ we solve (\ref%
{complex}) analytically for the trajectories 
\begin{equation}
x_{0}(t)=\pm a\arccos \left\{ \frac{\left[ (\kappa +\lambda )\cos \left( 
\frac{x_{0}}{a}\right) +\kappa -\lambda \right] e^{\frac{iht(\kappa +\lambda
)}{2a^{2}m}}+\lambda -\kappa }{\kappa +\lambda }\right\} .
\end{equation}%
For excited states one may easily solve these equations numerically as
depicted in figure \ref{F6}.

We observe the usual appearance of the only possible types of fixed points
in a Hamiltonian system, that is centres and saddle points. For the ground
state we observe a close resemblance of the qualitative behaviour with the
solution of the first excited state obtained for the harmonic oscillator as
shown in figure \ref{F2}.

\begin{figure}[ht]
\centering   \includegraphics[width=7.5cm,height=6.0cm]{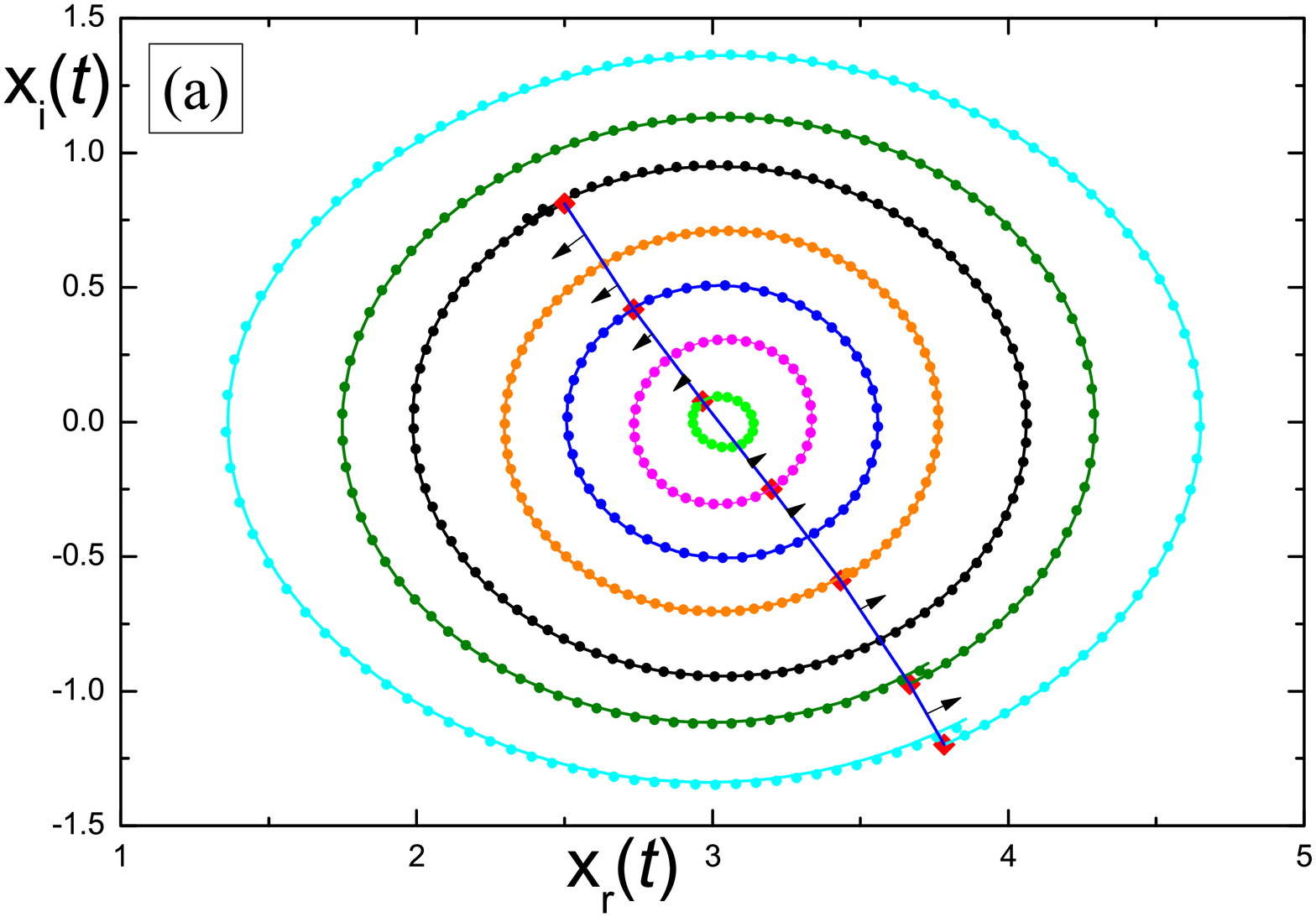} %
\includegraphics[width=7.5cm,height=6.0cm]{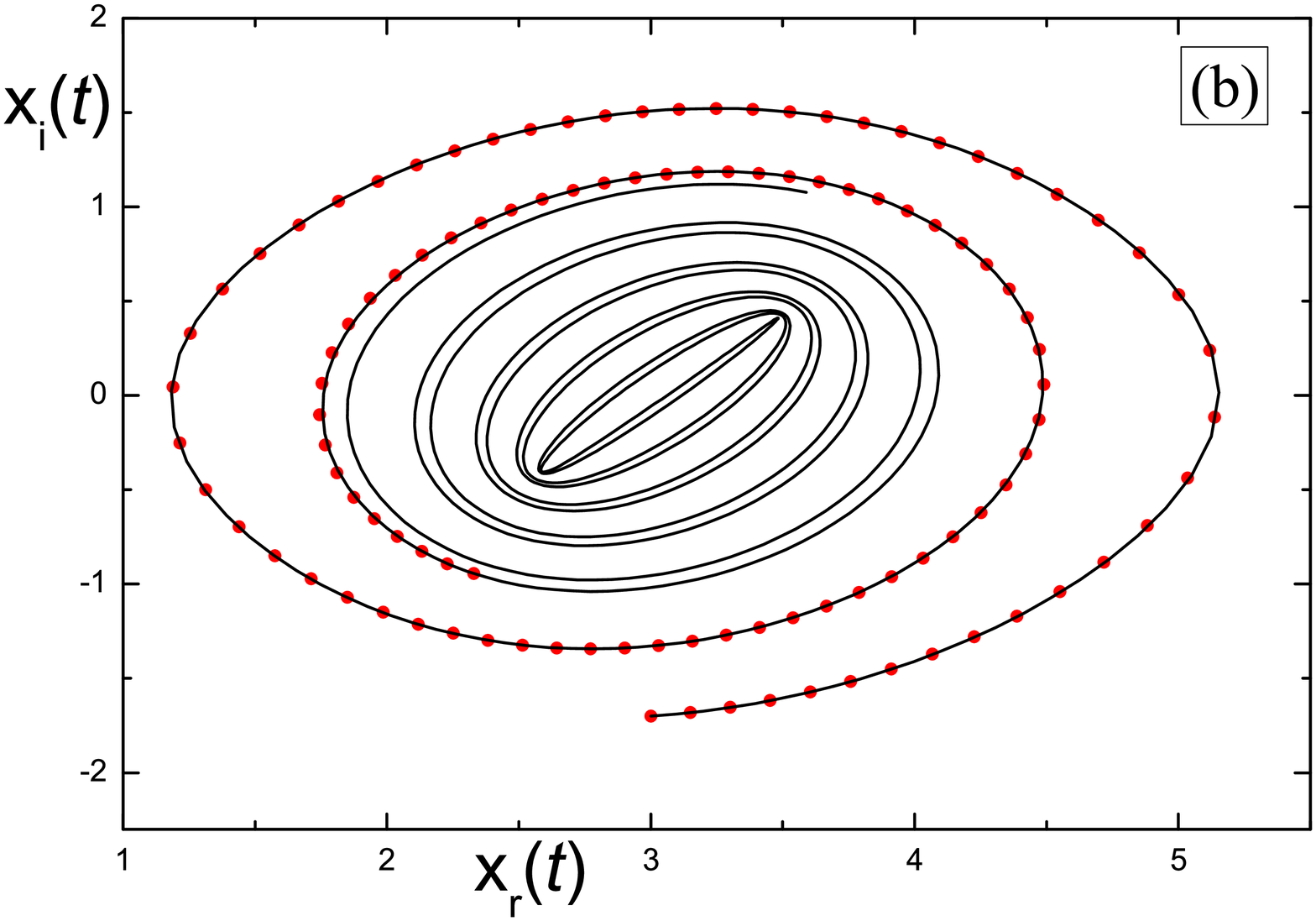}
\caption{Complex Bohmian trajectories as functions of time from Klauder
coherent states (scattered) versus classical trajectories corresponding to
solutions of (\protect\ref{H1}) and (\protect\ref{H2}) for the complex P\"{o}%
schl-Teller Hamiltonian $\mathcal{H}_{\text{PT}}$ (\protect\ref{PTcomplex})
(solid lines) for quasi-Poissonian distribution with $\protect\kappa =90$, $%
\protect\lambda =100$. (a) The evolution is shown from $t=0$ to $t=0.27$
with initial values on the isochrone with $t_{f}=0.04$ for $J=0.5$ and in
(b) from $t=0$ to $t=0.5$ (quantum) and $t=0$ to $t=3.0$ (classical) with
initial value $x_{0}=3-1.7i$ and $p_{0}=32.907+3.16416i$ for $J=20$. The
energy for this trajectory is therefore $E=\mathcal{H}_{\text{PT}%
}(x_{0},p_{0})=-8.44045+102.134i$.}
\label{F5}
\end{figure}

\begin{figure}[ht]
\centering   \includegraphics[width=7.5cm,height=6.0cm]{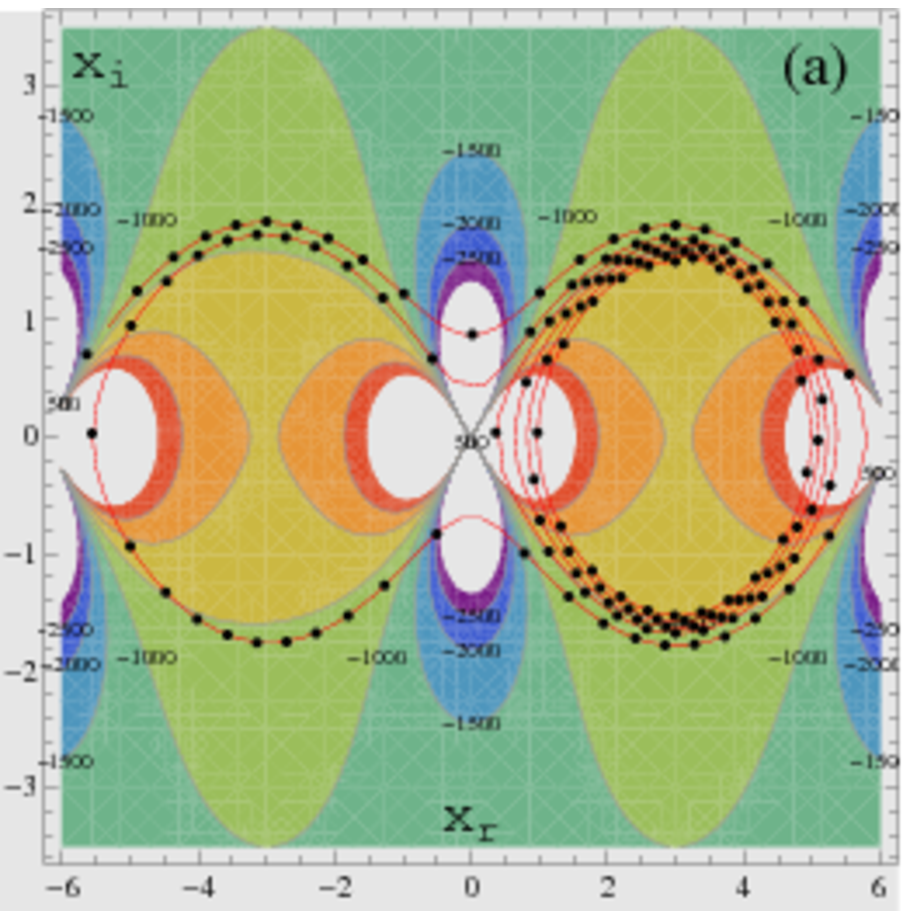}
\includegraphics[width=7.5cm,height=6.0cm]{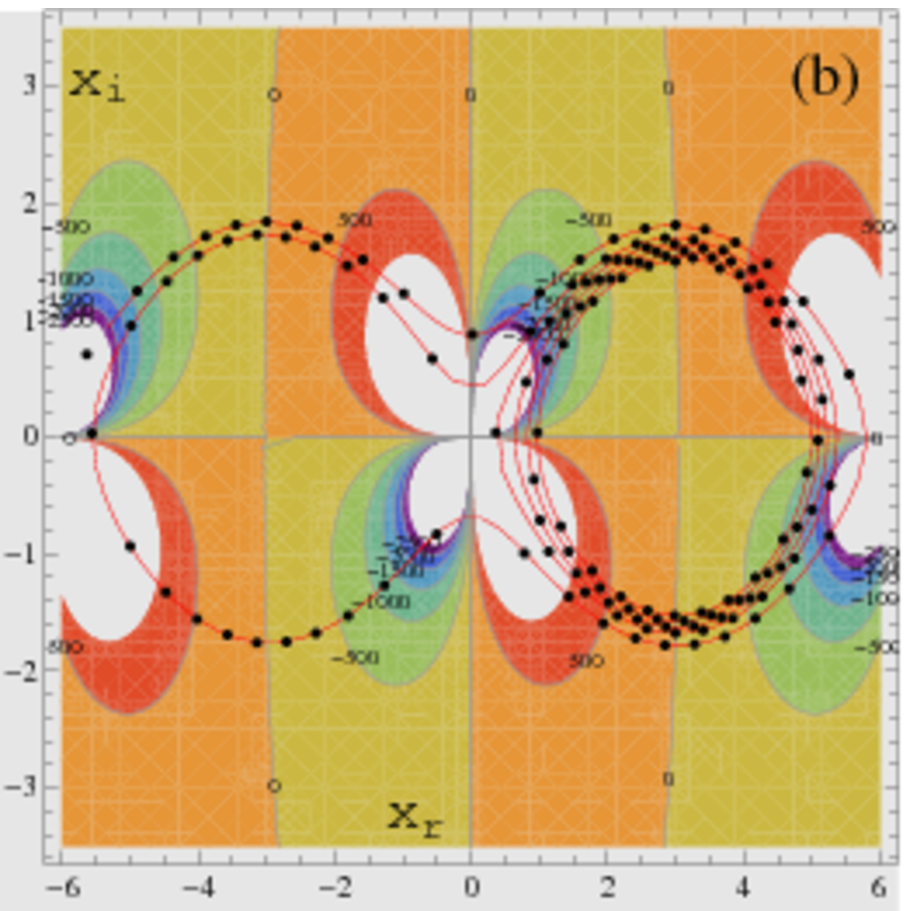}
\caption{Complex Bohmian quantum trajectories as functions of time from $%
J=0.5$-Klauder coherent states (scattered) versus classical trajectories
(solid lines) corresponding to solutions of (\protect\ref{H1}) and (\protect
\ref{H2}) for the complex P\"{o}schl-Teller potential $V_{\text{PT}}$ (a)
real part, (b) imaginary part for quasi-Poissonian and localized
distribution from $t=0$ to $t=1.78$ with $\protect\kappa =90$, $\protect%
\lambda =100$. The initial values are $x_{0}=3+1.5i$, $%
p_{0}=-30.1922+0.385121i$ such that $E=-6.55991-13.5182i$ (black solid, red
scattered) and $x_{0}=4.5$, $p_{0}=41.8376 i$ with real energy $E=-31.7564$
(blue solid, red scattered). }
\label{F11}
\end{figure}

\begin{figure}[h]
\centering   \includegraphics[width=7.5cm,height=6.0cm]{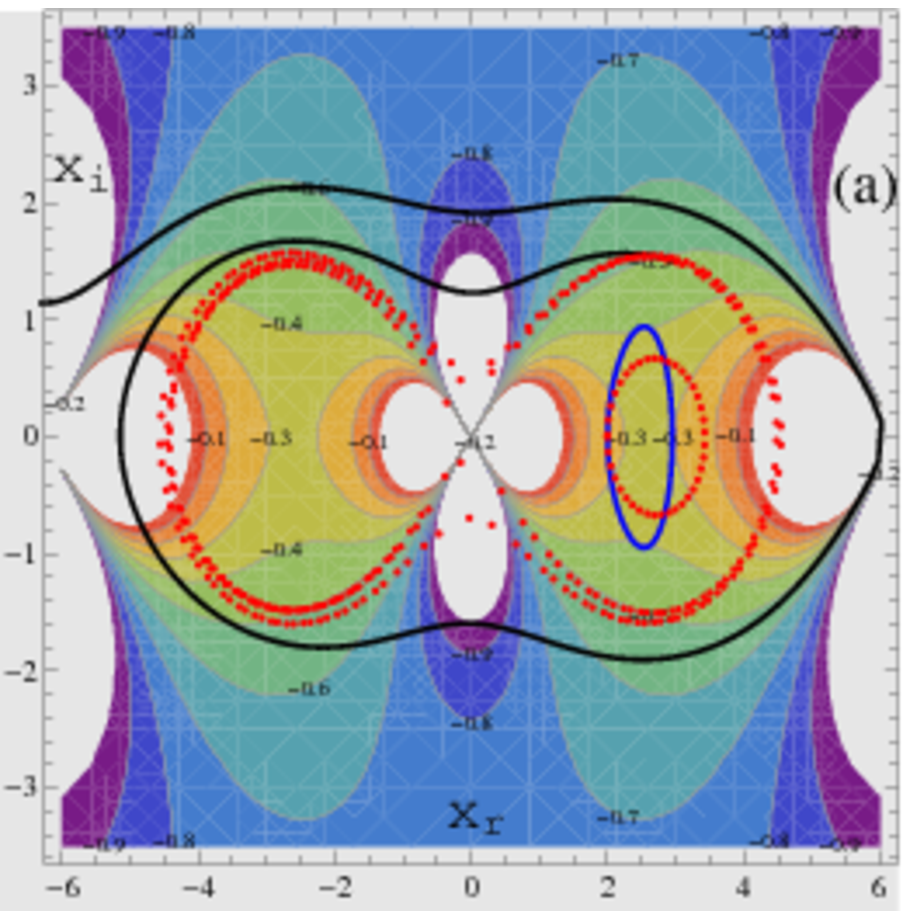} %
\includegraphics[width=7.5cm,height=6.0cm]{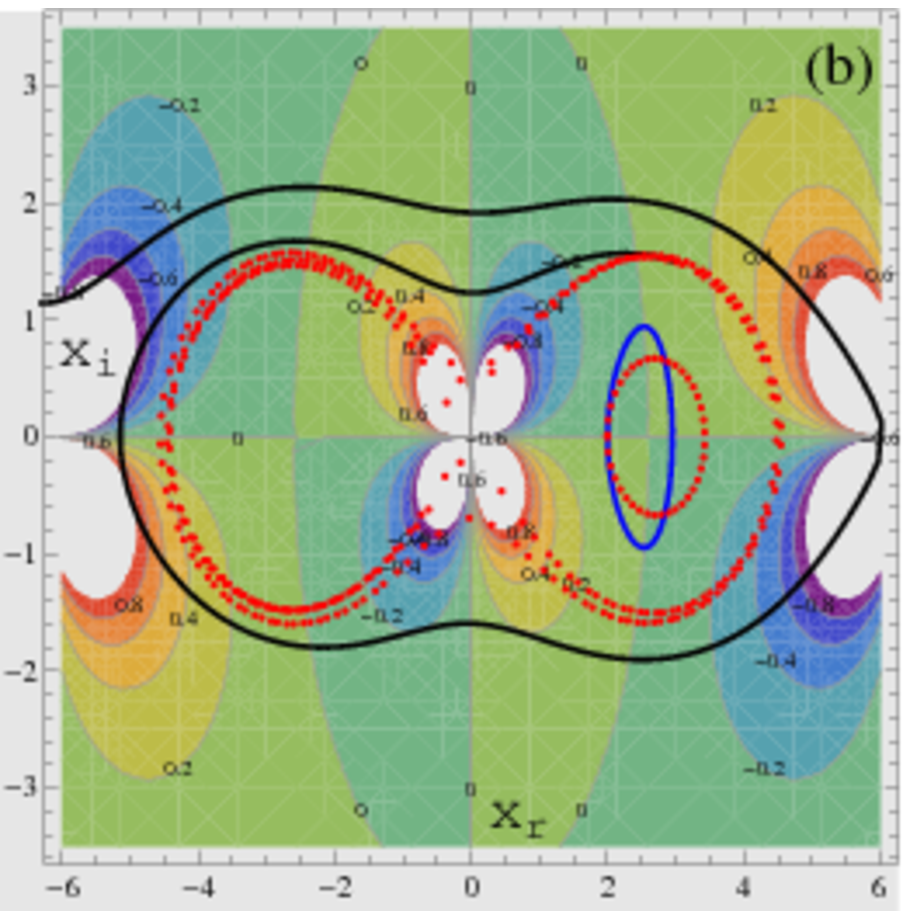}
\caption{Complex Bohmian trajectories as functions of time from $%
J=0.00057265 $-Klauder coherent states (scattered) versus classical (solid
lines) corresponding to solutions of (\protect\ref{H1}) and (\protect\ref{H2}%
) for the complex P\"{o}schl-Teller potential $V_{\text{PT}}$ (a) real part,
(b) imaginary part for the quasi-Poissonian regime and spead distribution
from $t=0$ to $t=100$ (quantum) and $t=0$ to $t=32$ (classical) with $%
\protect\kappa =2$, $\protect\lambda =3$. The initial values are $%
x_{0}=3+1.5i$ and $p_{0}=-0.788329+0.157336i$ such that $%
E=-0.187539-0.0275087i$ (black solid, red scattered) and $x_{0}=2$, $%
p_{0}=-0.49446 i$ with real energy $E=-0.277833$ (blue solid, red
scattered). }
\label{F12}
\end{figure}

Unlike the trajectories resulting from coherent states those obtained from
stationary states are not expected to have a similar behaviour to the purely
classical ones obtained from solving directly the equations of motion (\ref%
{H1}) and (\ref{H2}). Complexifying the variables as specified after (\ref%
{H1}) and (\ref{H2}), we may split the Hamiltonian into its real and
imaginary part $\mathcal{H}_{\text{PT}}=H_{r}+iH_{i}$ with%
\begin{eqnarray}
H_{r} &=&\frac{p_{r}^{2}-p_{i}^{2}}{2m}+V_{0}\left[ \frac{(\lambda
^{2}-\lambda )\left[ \cosh \left( \frac{x_{i}}{a}\right) \cos \left( \frac{%
x_{r}}{a}\right) +1\right] }{\left[ \cosh \left( \frac{x_{i}}{a}\right)
+\cos \left( \frac{x_{r}}{a}\right) \right] {}^{2}}-\frac{(\kappa
^{2}-\kappa )\left[ \cosh \left( \frac{x_{i}}{a}\right) \cos \left( \frac{%
x_{r}}{a}\right) -1\right] }{\left[ \cos \left( \frac{x_{r}}{a}\right)
-\cosh \left( \frac{x_{i}}{a}\right) \right] {}^{2}}\right]  \notag \\
&&-\frac{V_{0}}{2}(\lambda +\kappa )^{2},  \notag \\
H_{i} &=&\frac{p_{i}p_{r}}{m}+V_{0}\left[ \frac{(\lambda ^{2}-\lambda )\sinh
\left( \frac{x_{i}}{a}\right) \sin \left( \frac{x_{r}}{a}\right) }{\left[
\cosh \left( \frac{x_{i}}{a}\right) +\cos \left( \frac{x_{r}}{a}\right) %
\right] ^{2}}-\frac{(\kappa ^{2}-\kappa )\sinh \left( \frac{x_{i}}{a}\right)
\sin \left( \frac{x_{r}}{a}\right) }{\left[ \cos \left( \frac{x_{r}}{a}%
\right) -\cosh \left( \frac{x_{i}}{a}\right) \right] ^{2}}\right] .
\label{PTcomplex}
\end{eqnarray}%
This Hamiltonian also respects the aforementioned $\mathcal{PT}$-symmetry $%
\mathcal{PT}$: $x_{r}\rightarrow -x_{r}$, $x_{i}\rightarrow x_{i}$, $%
p_{r}\rightarrow p_{r}$, $p_{i}\rightarrow -p_{i}$, $i\rightarrow -i$.
Contourplots of the potential are shown in figures \ref{F11} and \ref{F12}
with the colourcode convention being associated to the spectrum of light
decreasing from red to violet. The corresponding equations of motion are
easily computed from (\ref{H1}) and (\ref{H2}), albeit not reported here as
they are very lengthy, and solved numerically as shown for some parameter
choices in the figures \ref{F5}, \ref{F11} and \ref{F12} as solid lines.

Let us now compare them with the complex Bohmian trajectories computed from
the Klauder coherent states (\ref{GK}). A previous initial attempt to
compute these trajectories has been made in \cite{MVJohnMathew}, however,
the preliminary computations presented there do not agree with our findings.
We start by depicting a case for the quasi-Poissonian distribution in figure %
\ref{F5}.

Remarkably, in that case we find a perfect match between these two entirely
different computations. We observe that unlike as for the real trajectories,
for which we required an effective potential to achieve agreement, these
computations are carried out in both cases for exactly the same coupling
constants $\kappa $ and $\lambda $ with no adjustments made. Thus, just as
for the harmonic oscillator, this suggests that the complex quantum
potential is simply a constant such that the effective potential essentially
coincides with the original one in (\ref{HPT}). From the trajectories with
larger radii in panel (a) we observe that the trajectories do not close and
are not perfect ellipses. Prolonging the time beyond the cut-off time in the
panel (a) scenario we find inwardly spiralling trajectories. The coincidence
between the purely classical calculation and the quantum trajectories still
persists for larger values of $J$ and more asymmetrical initial conditions
closer to the boundary of the potential as shown for an example in panel
(b). For larger values of time we encounter numerical problems due to the
poor convergence of the series for large values of $J$.

Our initial values in \ref{F5}(a) are chosen to lie on the isochrones, that
is the set of all points which when evolved in time will arrive all
simultaneously, say at $t_{f}$, on the real axis. The isochrone is indicated
in figure \ref{F5}(a) by a red line and an additional arrow attached to it
pointing in the direction in which the real axis is reached. In our example
the arrival time is chosen to be $t=0.04$. As discussed for instance in \cite%
{Goldfarb,ChouW} the wavefunction defined on the isoclines can be thought of
as leading to physical information as their corresponding complex quantum
trajectories acquire real values. Moreover, as shown in \cite{ChouW}, one
may even reconstruct the precise form of the entire wavefunction from the
knowledge of the isochrone and the information transported by the action. We
will follow up this line of enquiry elsewhere.

The qualitative behaviour of the classical trajectories can be understood by
considering the motion in the complex potential. We consider first a
trajectory of a particle with real energy depicted in figure \ref{F11} as
ellipse.

The initial position is taken to be on the real axis with the particle
getting a kick parallel to the imaginary axis. Within the real part of the
potential the particle starts on a higher potential level and would simply
role down further up into the upper half plane towards the imaginary axis
due to the curvature of the potential. However, the particle is also
subjected to the influence of the imaginary part of the potential and at
roughly $x_{r}=3$ this effect is felt when the particle reaches a turning
point, pulling it back to the real axis which is reached at the point when
reflecting $x_{i}$ at the turning point. At that point it has reached a
higher potential level from which it roles down back to the initial position
through the lower half plane in a motion similar to the one performed in the
upper half plane.

Trajectories with complex values for the initial values have in general also
complex energies. As can be seen in figure \ref{F11} for a specific case, we
obtain at first a qualitatively similar motion to the real case, with the
difference that the particle spirals outwards. In the real part of the
potential this has the effect that after a few turns the particle is
eventually attracted by the sink on top of the origin. The momentum it gains
through this effect propels it into the region with negative real part. Thus
the particle has bypassed the infinite potential barrier at the origin on
the real axis, tunneling to the next potential minimum, i.e. to the
forbidden region in the real scenario. Similar effects have been observed in
the purely classical treatment of a complex elliptic potential in \cite%
{Bender:2010eg}. The continuation of this trajectory and scenarios for other
parameter choices can be understood in a similar manner. For instance in
figure \ref{F12} we depict a trajectory which does not spiral at first, but
the particle has instead already enough momentum that allows it to tunnel
directly into the negative region.

As in the real scenario the Mandel parameter controls the overall
qualitative behaviour, although in the complex case this worsens for
non-real initial values that is complex energies. In quasi-Poissonian regime
pictured in figure \ref{F5} and \ref{F11} we observe a complete match
between the purely classical and the quantum computation. However, this
agreement ceases to exist in figure \ref{F12}, despite the fact that it is
showing a quasi-Poissonian case with the same value for $Q$. The difference
is that in the latter case the wavefunction is less well localized as we saw
in figure \ref{F7}. As can be seen in figure \ref{F12}, for real energies we
still have the same qualitative behaviour, but for complex energies the less
localized wave is spread across a wide range of potential levels such that
it can no longer mimic the same classical motion. However, qualitatively we
can see that in principle it is still compatible with the motion in a
complex classical P\"{o}schl-Teller potential.

\begin{figure}[h]
\centering   \includegraphics[width=7.5cm,height=6.0cm]{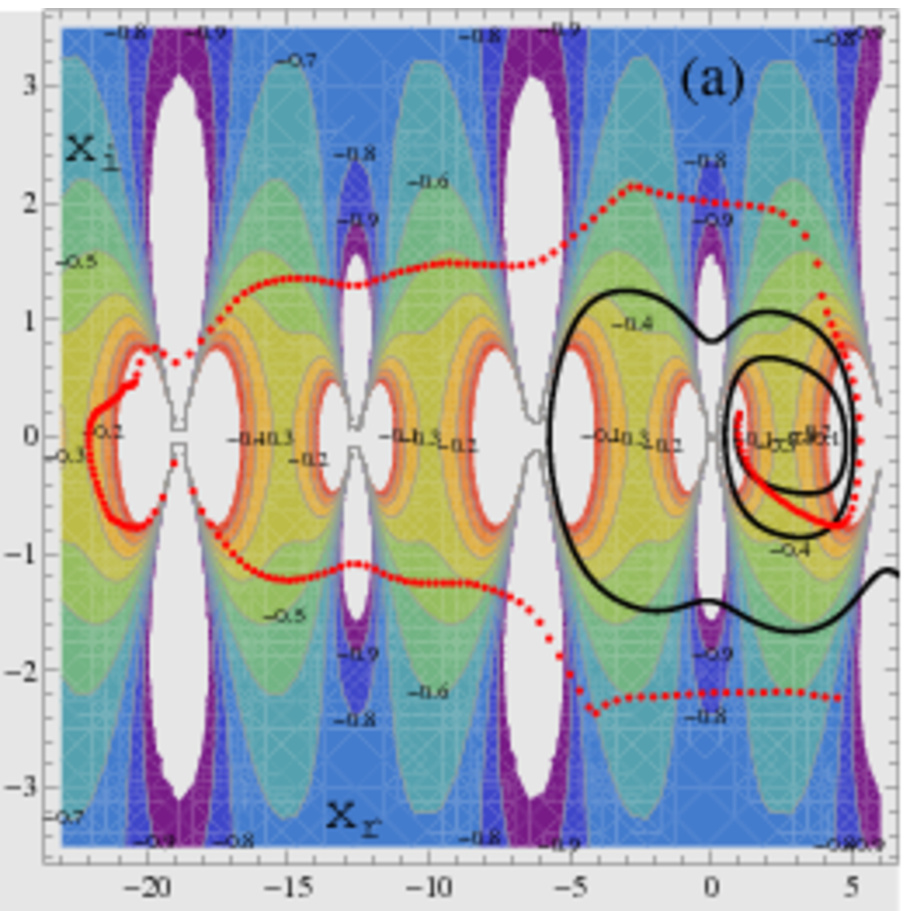} %
\includegraphics[width=7.5cm,height=6.0cm]{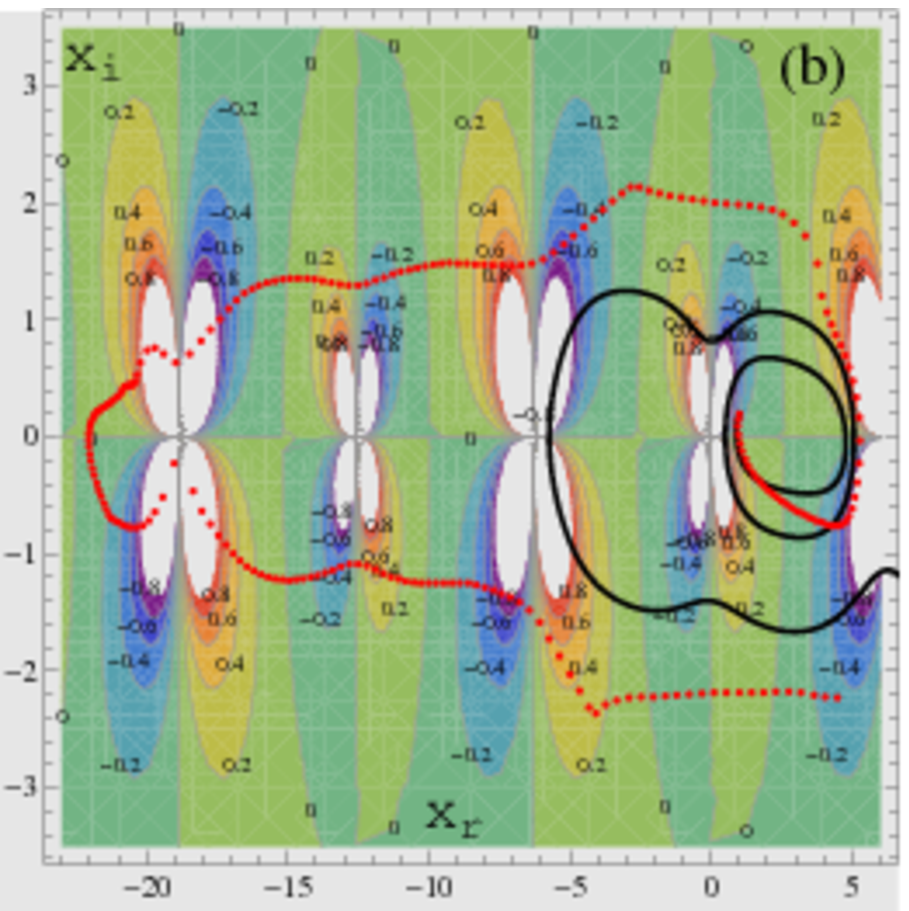}
\caption{Complex Bohmian trajectories as functions of time from $J=10 $%
-Klauder coherent states (scattered) versus classical (solid lines)
corresponding to solutions of (\protect\ref{H1}) and (\protect\ref{H2}) for
the complex P\"{o}schl-Teller potential $V_{\text{PT}}$ (a) real part, (b)
imaginary part for the sub-Poissonian regime from $t=0$ to $t=21$ (quantum)
and $t=0$ to $t=32$ (classical) with $\protect\kappa =2$, $\protect\lambda %
=3 $. The initial values are $x_{0}=2+0.2i$ and $p_{0}=-0.665052-0.406733i$
such that $E=-0.0941366-0.364635i$ (black solid, red scattered).}
\label{F13}
\end{figure}

As can be seen in figure \ref{F13} this resemblances ceases to exist when we
enter the sub-Poissonian regime.

We observe that the correlation between the two behaviours is now entirely
lost. Notably, the quantum trajectories enter into regions not accessible to
the classical ones. However, in a very coarse sense we can still explain the
overall behaviour of the quantum trajectories by appealing to the complex
potential.

\section{Conclusions}

We have computed real and complex quantum trajectories in two alternative
ways, either by solving the associated equation for the velocity or by
solving the Hamilton-Jacobi equations taking the quantum \ potential as a
starting point. In all cases considered we found perfect agreement for the
same initial values in the position.

Our main concern in this manuscript has been to investigate the quality of
the Klauder coherent states and test how close they can mimic a purely
classical description. This line of enquiry continues our previous
investigations \cite{DF2,DF5} for these type of states in a different
context. We have demonstrated that in the quasi-Poissonian regime well
localized Klauder coherent states produce the same qualitative behaviour as
a purely classical analysis. We found these features in the real as well as
in the complex scenario. For the real trajectories we conjectured some
analytical expressions reproducing the numerically obtained results. Whereas
the real case required always some adjustments, we found for the complex
analysis of the harmonic oscillator and the P\"{o}schl-Teller potential a
precise match with the purely classical treatment.

Naturally there are a number of open problems left: Clearly it would be
interesting to produce more sample computations for different types of
potentials, especially for the less well explored complex case. In that case
it would also be very interesting to explore further how the conventional
quantum mechanical description can be reproduced. Since Bohmian quantum
trajectories allow to establish that link there would be no need to guess
any rules in the classical picture mimicking some quantum behaviour as done
in the literature.

\medskip

\noindent \textbf{Acknowledgments:} SD is supported by a City University
Research Fellowship. AF thanks C. Figueira de Morisson Faria for useful
discussions.

\newif\ifabfull\abfulltrue

\end{document}